\documentclass{article}
\usepackage[utf8]{inputenc}
\usepackage{amsmath,amsfonts,amssymb,graphicx,epsfig,pdflscape,multirow,theorem,tikz,tikz-cd,braket}
\usepackage{subfigure}
\usepackage{fullpage}
\usepackage{geometry}
\usepackage{parskip}
\usepackage{bbm}
\usepackage{cleveref}
\usepackage{authblk}

\usepackage[maxbibnames=99,sorting=none]{biblatex} % bibnames stops "et al"
\DeclareFieldFormat{journaltitle}{#1}
\DeclareFieldFormat*{title}{\mkbibemph{#1\isdot}} %This command italicises the title of all bib items
\renewbibmacro{in:}{
  \ifentrytype{article}{
  }{
    \printtext{\bibstring{in}\intitlepunct}
  }
} % This disables the "in:" preceeding the Journal which is default behaviour....
\renewbibmacro*{volume+number+eid}{
  \printfield{volume}
  \setunit*{\addnbspace}
  \printfield{number}
  \setunit{\addcomma\space}
  \printfield{eid}}
\DeclareFieldFormat[article]{number}{\mkbibparens{#1}}

\DeclareFieldFormat{pages}{#1}
\renewbibmacro*{publisher+location+date}{
  \printlist{publisher}
  \setunit*{\addcomma\space}
  \printlist{location}
  \setunit*{\addcomma\space}
  \usebibmacro{date}
  \newunit}
\newcommand{\nc}{\newcommand}
\nc{\nn}{\nonumber\\ }
\nc{\chit}{\raisebox{0.25ex}{$\chi$}}
\nc{\chih}{\raisebox{0.25ex}{$\hat\chi$}}
\nc{\mfg}{\mathfrak{g}}
\nc{\mfh}{\mathfrak{h}}
\nc{\mfgh}{\widehat{\mathfrak{g}}}
\nc{\mfhh}{\widehat{\mathfrak{h}}}
\nc{\Ac}{\mathcal{A}}
\nc{\Bc}{\mathcal{B}}
\nc{\Acb}{\bar{\mathcal{A}}}
\nc{\Ic}{\mathcal{I}}
\nc{\Mc}{\mathcal{M}}
\nc{\Nc}{\mathcal{N}}
\nc{\Oc}{\mathcal{O}}
\nc{\Qc}{\mathcal{Q}}
\nc{\Vc}{\mathcal{V}}
\nc{\NS}{\mathrm{NS}}
\nc{\Vir}{\mathfrak{Vir}}
\newcommand{\asl}{\widehat{\mathfrak{sl}}}
\nc{\Virb}{\overline{\mathfrak{Vir}}}
\nc{\pa}{\partial}
\nc{\eps}{\epsilon}
\nc{\Gc}{\mathcal{G}}
\nc{\C}{\mathbb{C}}

\newcommand{\Z}{\mathbb{Z}}
\newcommand{\R}{\mathbb{R}}

\nc{\La}{\Lambda}
\nc{\Lh}{\hat{\Lambda}}
\nc{\I}{\mathbb{I}}
\nc{\ii}{\mathrm{i}}

\nc{\Hc}{\mathcal{H}}

\newtheorem{prop}{Proposition}[section]

\newcommand{\qed}{{\hfill$\blacksquare$\\}}
\newcommand{\prf}{{\underline{Proof:}\quad}}

\title{\Huge Asymmetric Galilean conformal algebras}
\date{}
\author[1]{\Large Eric Ragoucy}
\author[2]{J{\o}rgen Rasmussen}
\author[3]{Christopher Raymond}
\affil[1]{Laboratoire d'Annecy-le-Vieux de Physique Th\'eorique (LAPTH) \protect\\
Universit\'e Grenoble Alpes, CNRS, F-74000 Annecy, France \protect\\
{\tt ragoucy\,@\,lapth.cnrs.fr}\protect\\[-.1cm]
\mbox{}}
\affil[2]{School of Mathematics and Physics, the University of Queensland \protect\\
St Lucia, Brisbane, Queensland 4072, Australia \protect\\
{\tt j.rasmussen\,@\,uq.edu.au}\protect\\[-.1cm]
\mbox{}}
\affil[3]{Mathematical Sciences Institute, The Australian National University \protect\\
Acton, Australian Capital Territory 2600, Australia \protect\\
{\tt christopher.raymond\,@\,anu.edu.au}}
\begin{document}

\maketitle
\begin{abstract}
The usual Galilean contraction procedure for generating new conformal symmetry algebras takes as input a number of symmetry algebras which are equivalent up to central charge. We demonstrate that the equivalence condition can be relaxed by inhomogeneously contracting the chiral algebras and present general results for the ensuing asymmetric Galilean algebras. Several examples relevant to conformal field theory are discussed in detail, including superconformal algebras and W-algebras. We also discuss how the Sugawara construction is modified in the asymmetric setting.
\end{abstract}

\section{Introduction}
Galilean conformal algebras provide important families of conformal field theories (CFTs) \cite{BG09,BGMM10}. They arise as symmetry algebras of toy gravity models relevant to holography \cite{GG11,Vas12,BF12,ABFGR13,BO14} (see \cite{Obl17} for a review), and in the study of tensionless strings \cite{Bag13,BBCP18,BBCP20,BBCDP20}. They also encode the symmetries of a two-dimensional non-relativistic conformal system \cite{Hag72,DH09,BGMM10,BM10,HR10}.

Galilean symmetry algebras can be constructed in several ways. In this paper, we focus on a parametric contraction procedure known as Galilean contraction, analogous to an {\'I}n{\"o}n{\"u}-Wigner contraction of Lie algebras \cite{IW53,Sale61}. Galilean algebras can also be constructed as so-called Takiff algebras, as in \cite{Tak71,BR12,RRR20,Que20}, or by using the so-called semigroup expansion method, as in \cite{IRS06,CCRS18,CIRR20}. Moreover, they appear naturally as the flat-space limit symmetries of models on $AdS_{3}$ spacetime \cite{BH86,BC07,BL13,GMPT13,CF17}.

The Galilean contraction procedure, as studied in \cite{BG09,BGMM10,GRR14,RR17}, takes as input two symmetry algebras which are equivalent up to central parameters. That is, the chiral algebras are the same operator product algebra (OPA) \cite{Thielemans,RR17}, up to the value of their central charge. The procedure involves a parameter-dependent ($\eps$) map on a particular set of fields such that in the limit $\eps \to 0$ the map becomes singular. If the limit is well-defined, the resulting algebra is referred to as the Galilean algebra corresponding to the input algebras. The most studied example of this construction involves contracting two copies of the Virasoro algebra, producing the so-called Galilean conformal algebra, also known as the Galilean Virasoro algebra \cite{DZ07,BGMM10,RR17}.

The Galilean contraction procedure has been generalised to allow for the input of any number of symmetry algebras, leading to so-called higher-order Galilean contractions \cite{CCRS18,RR19}. The ensuing Galilean algebras exhibit graded structures of any integer order $N\geqslant2$. In \cite{RRR20}, the contraction procedure was further generalised to accommodate a tensor-product structure. This results in the so-called multi-graded Galilean contraction and produces Galilean algebras graded by integer sequences.

Here, we relax the condition on the equivalence of input algebras by inhomogeneously rescaling the fields generating the algebras. We refer to this as {\em asymmetric Galilean contraction}. 

We require that each input algebra admits an embedding of a particular algebra, meaning that the input algebras have a pair of subalgebras which are equivalent up to the value of their central parameters. The fields of these subalgebras are contracted according to the usual Galilean procedure, yielding an order-two Galilean subalgebra. The fields not in the equivalent subalgebras are rescaled separately, and the choice of these rescalings determines the structure of the asymmetrically contracted algebra arising in the limit.

Algebras which can be realised using asymmetric Galilean contractions have appeared in the literature before, and include WZW models on non-semisimple Lie groups \cite{Moham94,ORS94,Sfet94,HKOC96}, supersymmetric tensionless strings \cite{Bag13, BBCP18}, and super-BMS algebras \cite{LM16,BJLMN16,BDR17,FMT17,BLN17,CCFR18,BBLN18,BBNN19}. In fact, several of these papers employ inhomogeneous contraction techniques which are special cases of the general construction presented here.

The paper is structured as follows. We begin by presenting a general description of asymmetric Galilean contractions. We then provide a number of detailed examples chosen to demonstrate the procedure and to provide interesting physical applications. Finally, we discuss a Sugawara construction for asymmetrically contracted affine Lie algebras.

%*********************************************
\section{Review of operator product algebras}
%*********************************************
The key structure in this paper is the symmetry algebra of a CFT. Such an algebra can be realised as an OPA, here denoted by $\Ac$, which is particularly convenient for studying contractions. We note, however, that asymmetric contractions and our results apply more broadly, such as to finite-dimensional Lie algebras and other algebraic structures encoding symmetries.

The fundamental objects of an OPA are the fields $A(z) \in \Ac$, which have mode expansions
\begin{equation}
    A(z) = \sum_{n \in \Z - \Delta_{A}} A_{n}z^{-n-\Delta_{A}},
\end{equation}
where $\Delta_{A} \in \R$ is known as the conformal weight of the field $A$. The basic product in an OPA is the operator product expansion (OPE). The OPE between two fields $A(z),B(w) \in \Ac$ is given by
\begin{equation}\label{OPE}
    A(z)B(w) \sim \sum_{n=1}^{\Delta_{A} + \Delta_{B}} \frac{[AB]_{n}(w)}{(z-w)^{n}},
\end{equation}
where $[AB]_{n}$ is a field of conformal weight $\Delta_{A} + \Delta_{B} - n$. The symbol $\sim$ indicates that all non-singular terms are omitted. A set of fields is said to {\em generate} an OPA if all other fields in the OPA can be obtained from the generating set by taking OPEs, linear combinations, normally-ordered products, and derivatives.

An OPA is said to be {\em conformal} if it contains a distinguished field $T(z)$ that generates a Virasoro subalgebra:
\begin{equation}
    T(z)T(w) \sim \frac{c/2}{(z-w)^4} + \frac{2T(w)}{(z-w)^2} + \frac{\partial T(w)}{z-w}.
\end{equation}
A field $A(z) \in \Ac$ is called a {\em scaling field} if
\begin{equation}
    [TA]_{2} = \Delta_{A}A, \qquad [TA]_{1} = \partial A.
\end{equation}
A scaling field $A(z)$ is said to be {\em quasi-primary} if $[TA]_{3} = 0$, and {\em primary} if $[TA]_{n} = 0$ for all $n > 2$. Primary fields are thus quasi-primary, and we note that the Virasoro field $T(z)$ is quasi-primary.

Let $\mathcal{Q}$ denote a basis for the space of quasi-primary fields of a conformal OPA $\Ac$. The OPE between two fields in $\mathcal{Q}$ is given by
\begin{equation}\label{QPOPE}
    A(z)B(w) \sim \sum_{Q \in \mathcal{Q}}f^{A}{}^{B}_{Q}\left( \sum_{n=0}^{\Delta_{A} + \Delta_{B} - \Delta_{Q}} \frac{\beta_{\Delta_{Q};n}^{\Delta_{A},\Delta_{B}}\partial^{n} Q(w)}{(z-w)^{\Delta_{A} + \Delta_{B} - \Delta_{Q} - n}} \right),
\end{equation}
where $f^{A}{}^{B}_{C}$ are structure constants, and
\begin{equation}
    \beta_{\Delta_{Q};n}^{\Delta_{A},\Delta_{B}} = \frac{(\Delta_{A}- \Delta_{B}+\Delta_{Q})_{n}}{n!(2\Delta_{Q})_{n}}, \qquad (x)_{n} = \prod_{j=0}^{n-1}(x+j).
\end{equation}
The expression in parentheses in \eqref{QPOPE} is known as the conformal chain (in reference to similar quantities appearing in \cite{Zam85}), and is determined by $\Delta_{A}, \Delta_{B},$ and $Q$. 
Using this observation, the explicit dependence of the fields on the variables $z,w$ will be dropped in the following, and we will write the OPE 
\eqref{QPOPE} as
\begin{equation}
    A \times B \simeq \sum_{Q \in \mathcal{Q}}f^{A}{}^{B}_{Q} \{ Q \},
\end{equation}
where we have used $\times$ and $\simeq$ to distinguish this form of the product from that of \eqref{QPOPE}.

An important family of OPAs are those whose underlying algebra of modes is an infinite-dimensional Lie algebra. Such OPAs are known as {\em Lie-type} OPAs. Usually, we consider symmetries described by infinite-dimensional Lie algebras with central extensions. As we are working with the field algebras associated with these infinite-dimensional Lie algebras, the value that the central extension takes on the vacuum representation becomes a parameter of the theory. We will refer to this parameter as a \emph{central parameter} and denote it by $c \in \mathbb{C}$ generally, although it may be, for example, the level $k$ of an affine current algebra. 

By definition, only the structure constants accompanying the identity field depend on central parameters for a Lie-type algebra. Moreover, the structure constants accompanying the identity are linear in $c$. Hence, the OPE of a Lie-type algebra may be written as
\begin{equation}
    A \times B \simeq c \, f^{A}{}^{B}_{\I} \{ \I \} + \sum_{C \in \bar{\mathcal{Q}}} f^{A}{}^{B}_{C} \{ C \},
\end{equation}
where $\I$ denotes the identity field of the OPA. The remaining terms are a sum over $\bar{\mathcal{Q}}\equiv \mathcal{Q} \setminus \{ \I \}$. Later in the paper, we will employ the following summation convention: When the identity field is exhibited explicitly on the right-hand side, accompanying summations are taken to be over all fields excluding the identity. 

A set of {\em elementary} fields of an OPA are those fields from which all other fields can be constructed by taking linear combinations, normally-ordered products, and derivatives. Note that this definition does not include taking OPEs. While the set of elementary fields does generate the OPA, it is not necessarily a minimal generating set. For a Lie-type OPA, the modes of elementary fields span the underlying infinite-dimensional Lie algebra, with central parameters replaced by central elements. The singular part of the OPE between elementary fields of a Lie-type OPA features only elementary fields and their derivatives, along with the corresponding structure constants. 

The OPAs associated with infinite-dimensional Lie algebras, such as affine Lie algebras, the Virasoro algebra, and superconformal algebras, provide interesting examples of Lie-type OPAs. Further details on the algebraic structure of OPAs is given in \cite{Thielemans,RR17}.

%*********************************************
\section{Asymmetric Galilean algebras}
\label{Sec:GenFramework}
%*********************************************
We begin by considering a Lie-type OPA $\Hc$, and denote a corresponding set of elementary fields by $H$. We then consider embeddings of $\Hc$ into Lie-type OPAs $\Ac_{1}$ and $\Ac_{2}$, where $\Ac_{1}$ and $\Ac_{2}$ are not necessarily equivalent up to central parameters. We denote the image of $\Hc$ under the embedding into $\Ac_{\ell}$ by $\Hc_{\ell}$, $\ell = 1,2$. The central parameter of the algebra $\Ac_{\ell}$ is denoted by $c_{(\ell)}$. In the corresponding infinite-dimensional Lie algebras, where one has central elements rather than central parameters, the subalgebras $\Hc_{\ell}$ are indeed isomorphic.

For each $\ell$, we can extend the set of elementary fields of $\Hc_{\ell}$, denoted by $H_{\ell}$, to a set of elementary fields for $\Ac_{\ell}$, denoted by $G_{\ell}$. We thus have the partitions $G_{\ell} = H_{\ell} \sqcup \bar{G}_{\ell}$. For simplicity, we will assume that the elementary fields of $\Hc_{1}$ and $\Hc_{2}$ have been selected such that (loosely speaking) $H_{1} = H_{2}$. We denote the vector subspace of $\Ac_{\ell}$ spanned by the fields in $G_{\ell},H_{\ell}$, and $\bar{G}_{\ell}$, along with their derivatives, by $\mfg_{\ell}, \mfh_{\ell},$ and $\bar{\mfg}_{\ell}$, respectively.

We will use notation $A_{(\ell)}, B_{(\ell)}, C_{(\ell)}$ for fields in $H_{\ell}$, and $X_{(\ell)}, Y_{(\ell)}, Z_{(\ell)}$ for fields in $\bar{G}_{\ell}$. Moreover, we generally denote the structure constants of $\Ac_1$ and $\Ac_2$ by $f_{(1)}$ and $f_{(2)}$, respectively, but will denote the structure constants of $\Hc$ simply by $f^{A}{}^{B}_{C}$. 

We remark that, although we assume in the following that the input algebras are Lie-type OPAs, it is possible to perform asymmetric contractions on non-Lie-type algebras. Examples of asymmetric Galilean contractions of non-Lie-type $W$-algebras are thus explored in Section \ref{Sec:AsymExamplesDouble}.

A Galilean-type contraction on the OPA
\begin{equation}
    \Ac = \Ac_1\otimes \Ac_2
\end{equation}
begins by choosing a map on the set of elementary fields $G_1 \cup G_2$, which is parametrized by $\eps \in \mathbb{C}$. For each pair $m_1,m_2\in \R_{\geq 0}$, we thus introduce new fields
\begin{equation}
    A_{i,\eps} = \eps^{i}\left( A_{(1)} + (-1)^{i}A_{(2)} \right), \qquad i \in \{ 0,1 \},
\end{equation}
and
\begin{equation}
    X_{1,\eps} = \eps^{m_1} X_{(1)}, \qquad X_{2,\eps} = \eps^{m_2} X_{(2)},
\end{equation}
along with new central parameters
\begin{equation}
    c_{i,\eps} = \eps^{i}\left( c_{(1)} + (-1)^{i}c_{(2)} \right), \qquad i \in \{ 0,1 \}.
\end{equation}
The inverse map on fields is given by
\begin{equation}
    A_{(1)} = \frac{1}{2}\left( A_{0,\eps} + \eps^{-1}A_{1,\eps} \right), \quad A_{(2)} = \frac{1}{2} \left( A_{0,\eps} - \eps^{-1}A_{1,\eps} \right), \quad 
    X_{(1)} = \eps^{-m_1}X_{1,\eps}, \quad X_{(2)} = \eps^{-m_2}X_{2,\eps},
    \label{asyminvbasis}
\end{equation}
and correspondingly, we have an inverse map on central parameters, given by
\begin{equation}
    c_{(1)} = \frac{1}{2}\left( c_{0,\eps} + \eps^{-1}c_{1,\eps} \right), \quad c_{(2)} = \frac{1}{2} \left( c_{0,\eps} - \eps^{-1}c_{1,\eps} \right).
\end{equation}
For simplicity, for each $\ell$, all elements $X_{(\ell)} \in \bar{G}_\ell$ are scaled by the {\em same} $\eps$-monomial. We will comment on this assumption at the end of this Section.

The contraction is now performed by taking the limit $\eps \to 0$. If the limit exists, it defines a map $\Ac \to \Ac_{G}$ to the corresponding {\em asymmetric Galilean algebra}, where
\begin{equation}\label{galmap}
    A_{i,\eps} \mapsto A_{i}, \qquad X_{1,\eps} \mapsto X_{1}, \qquad X_{2,\eps} \mapsto X_{2}.
\end{equation}
The images of the elementary fields of the OPA $\Ac$ form a set of elementary fields of $\Ac_{G}$.

We denote the asymmetric Galilean contraction of $\Ac_1$ and $\Ac_2$ with respect to particular embeddings $\rho_{\ell}$ of the subalgebra $\Hc$ by $\Ac_{G} = \left( \Ac_1^{m_1} \overset{\rho_{1}}{\hookleftarrow} \Hc \overset{\rho_{2}}{\hookrightarrow} \Ac_2^{m_2} \right)_{G}$, or simply by $\left( \Ac_1^{m_1} \hookleftarrow \Hc \hookrightarrow \Ac_2^{m_2} \right)_{G}$. By construction, there is an exchange symmetry in the asymmetric Galilean contracted algebras given by
\begin{equation}
\left( \Ac_1^{m_1} \overset{\rho_{1}}{\hookleftarrow} \Hc \overset{\rho_{2}}{\hookrightarrow} \Ac_2^{m_2} \right)_{G}\equiv  \left( \Ac_2^{m_2} \overset{\rho_{2}}{\hookleftarrow} \Hc \overset{\rho_{1}}{\hookrightarrow} \Ac_1^{m_1} \right)_{G}.
\end{equation}

The contracted algebra is graded according to the $\eps$-monomials used in the procedure. Concretely, we have that the set of elementary fields of $\Ac_{G}$ splits as $H^0 \sqcup H^{1} \sqcup \bar{G}^{m_{1}}_{1}\sqcup \bar{G}^{m_{2}}_{2}$, where the superscript denotes the Galilean grading of the fields. We denote the vector subspaces of $\Ac_{G}$ formed by the fields in $H^0, H^{1}, \bar{G}^{m_{1}}_{1},$ and $\bar{G}^{m_{2}}_{2}$, along with their derivatives, by $\mfh^{0}, \ \mfh^{1}, \ \bar{\mfg}^{m_1}_{1},$ and $\bar{\mfg}^{m_2}_{2}$, respectively. The resulting algebra $\Ac_{G}$ has an order-two Galilean subalgebra, denoted by $\Hc_{G}$, with $H_{G} = H^{0} \sqcup H^{1}$ as a set of elementary fields.

In general, for Lie-type OPAs, products between elementary fields are sufficient to determine the algebraic structure. As such, we use notation $\mfh^{i} \times \mfh^{j}$ to denote the space of OPEs between fields in $\mfh^{i}$ and $\mfh^{j}$, $i,j \in \{ 0,1 \}$. For example, we write
\begin{equation}
    \mfh^{0} \times \mfh^{1} \subseteq \mfh^{1},
\end{equation}
to describe that OPEs between elementary fields in $H^{0}$ and $H^{1}$, and their derivatives, only produce fields in $\mfh^{1}$.

Naturally, the structure of $\Ac_{G}$ is sensitive to the choice of the contraction parameters $m_\ell$; indeed, $\Ac_{G}$ may only exist for certain choices. A simplified situation occurs if $\Ac_2 \cong \Hc_2$. In that case, there is only one contraction parameter as $\bar{G}_2$ is empty, and we denote the resulting Galilean algebra simply by $\left(\Hc \hookrightarrow \Ac^{n} \right)_{G}$.

\begin{prop}
\label{mainprop}
Given Lie-type OPAs $\Ac_{1}$, $\Ac_{2}$, and $\Hc$; embeddings $\Hc \overset{\rho_{1}}{\hookrightarrow} \Ac_{1}$ and $\Hc \overset{\rho_{2}}{\hookrightarrow} \Ac_{2}$; and choices of parameters $m_{1},m_{2} \in \mathbb{R}_{\geq 0}$; consider the asymmetric Galilean OPA 
\begin{equation*}
\Ac_{G} = \left( \Ac_1^{m_1} \hookleftarrow \Hc \hookrightarrow \Ac_2^{m_2} \right)_{G}. 
\end{equation*}

The algebraic structure for the subalgebra $\Hc_{G}$ is given by
$$
\mfh^0\times \mfh^j \subseteq \mfh^j,\ j=0,1 ;\qquad \mfh^1\times \mfh^1\simeq \{0\};
$$

while $\bar{\mfg}^{m_1}_1 \times \bar{\mfg}^{m_2}_2\simeq \{ 0 \}$ for all $m_1,m_2$. 

The remaining algebraic structure depends on the choice of $m_1$ and $m_2$. The OPEs involving fields in $\bar{\mfg}_{\ell}^{m_{\ell}}$, $\ell = 1,2$, are determined by the value of $m_{\ell}$. The OPEs for $\ell = 1,2$, are independent of each other and fall into the following cases:

\begin{enumerate}
\item If $m_\ell=0$, which is possible only if $\mfg_{\ell} \times \bar{\mfg}_{\ell} \subseteq \bar{\mfg}_{\ell}$, then
$$ 
\mfh^0\times \bar{\mfg}_{\ell}^{0}\subseteq \bar{\mfg}_{\ell}^{0}\ ;\qquad 
\mfh^1\times \bar{\mfg}_{\ell}^{0}\simeq \{0\}\ ;\qquad 
\bar{\mfg}_{j}^{0}\times \bar{\mfg}_{\ell}^{0} \subseteq \bar{\mfg}_{\ell}^{0}\,.
$$
\item If $m_\ell =\frac12$, which is possible only if 
$\mfh_{\ell} \times \bar{\mfg}_{\ell} \subseteq \bar\mfg_{\ell}$ in the original algebra $\Ac_{\ell}$, then
$$ 
\mfh^0\times \bar{\mfg}_{\ell}^{\frac12}\subseteq \bar{\mfg}_{\ell}^{\frac12}\ ;\qquad 
\mfh^1\times \bar\mfg_{\ell}^{\frac12}\simeq \{0\}\ ;\qquad 
\bar{\mfg}_{\ell}^{\frac12}\times \bar{\mfg}_{\ell}^{\frac12} \subseteq \mfh^{1}\,.
$$
\item If $m_\ell=1$, which is always possible, then
$$ 
\mfh^0\times \bar{\mfg}_{\ell}^{1}\subseteq \bar\mfg_{\ell}^{1}\oplus \mfh^1\ ;\qquad 
\mfh^1\times \bar{\mfh}_{\ell}^{1}\simeq \{0\}\ ;\qquad
\bar{\mfg}_{\ell}^{1}\times \bar{\mfg}_{\ell}^{1} \simeq \{0\}\,.
$$
\item If $m_\ell\notin \{ 0, \frac{1}{2},1 \}$, and the contraction limit is well-defined, then
$$
\mfh^{0} \times \bar{\mfg}^{m_\ell}_{\ell} \subseteq \bar{\mfg}^{m_\ell}_{\ell}; \qquad \mfh^{1} \times \bar{\mfg}^{m_\ell}_{\ell} \simeq \{ 0 \}; \qquad \bar{\mfg}^{m_\ell}_{\ell} \times \bar{\mfg}^{m_\ell}_{\ell} \simeq \{ 0 \}.
$$
\end{enumerate}
\end{prop}
\prf
To determine the structure of the contracted algebra, we begin by calculating OPEs in the algebra $\Ac$ before taking the contraction limit. As the algebras $\Ac_{1}$ and $\Ac_{2}$ are Lie-type, their structure constants accompanying the identity are linear in $c$. Recalling that $c_{i,\eps} = \eps^{i}(c_{(0)} + (-1)^ic_{(1)})$, where $i \in \{ 0,1 \}$, the OPEs between elementary fields in the subalgebras $\Hc_{\ell}$ are given by
\begin{equation}
A_{i,\eps} \times B_{j,\eps} \simeq
       c_{i+j,\eps}\,f^{A}{}^{B}_{\I}\{\I\} + f^{A}{}^{B}_{C} \ \{ C_{i+j,\eps} \}, \quad \mathrm{for} \ i+j \leq 1,
    \label{preconAB}
\end{equation}
where $i,j \in \{ 0,1 \}$, and we have that $A_{1,\eps} \times B_{1,\eps} = \eps^2 A_{0,\eps} \times B_{0,\eps}$.

The remaining products between transformed elementary fields of $\Ac$ are given by
\begin{equation}\begin{aligned}
A_{i,\eps} \times X_{1,\eps} &\simeq \left[ \eps^{i+m_1} \frac{c_{0}\,f_{(1)\, \I}^{AX}}{2} \{ \I \} + \eps^{i+m_1-1}\frac{c_{1}\,f_{(1)\, \I}^{AX}}{2} \{ \I \} + \eps^{i} f_{(1)\, Y}^{AX} \{ Y_{1,\eps} \} \right. \\
& \ \ \left. + \eps^{i+m_1} \frac{f_{(1)\,B}^{AX}}{2} \{ B_{0,\eps} \}+ \eps^{i+m_1-1} \frac{f_{(1)\, B}^{AX}}{2} \{ B_{1,\eps} \} \right],
\label{preconAX1}
\end{aligned}\end{equation}
\begin{equation}\begin{aligned}
A_{i,\eps} \times X_{2,\eps} &\simeq(-1)^{i} \left[ \eps^{i+m_2} \frac{c_0\,f_{(2)\, \I}^{AX}}{2} \{ \I \} - \eps^{i+m_2-1}\frac{c_1\,f_{(2)\, \I}^{AX}}{2} \{ \I \} + \eps^{i} f_{(2)\, Y}^{AX} \{ Y_{2,\eps} \} \right. \\
& \ \ \left. + \eps^{i+m_2} \frac{f_{(2)\, B}^{AX}}{2} \{ B_{0,\eps} \}- \eps^{i+m_2-1} \frac{f_{(2)\, B}^{AX}}{2} \{ B_{1,\eps} \} \right],
\label{preconAX2}
\end{aligned}\end{equation}
\begin{equation}\begin{aligned}
X_{1,\eps} \times Y_{1,\eps} &\simeq \left[ \eps^{2m_1}\frac{c_0\,f_{(1)\, \I}^{XY}}{2} \{ \I \} + \eps^{2m_1-1}\frac{c_1 \, f_{(1)\, \I}^{XY}}{2} \{ \I \}  + \eps^{m_1} f_{(1)\, Z}^{XY} \{ Z_{1,\eps} \}\right.\\
& \ \ +\left. \eps^{2m_1}\frac{f_{(1)\,A}^{XY}}{2}\{ A_{0,\eps} \} + \eps^{2m_1-1} \frac{f_{(1)\, A}^{XY}}{2} \{ A_{1,\eps} \} \right],
\label{preconXY1}
\end{aligned}\end{equation}
and
\begin{equation}\begin{aligned}
X_{2,\eps} \times Y_{2,\eps} &\simeq \left[ \eps^{2m_2}\frac{c_0 \, f_{(2)\, \I}^{XY}}{2}\{ \I \} - \eps^{2m_2-1}\frac{c_1 \, f_{(2)\,\I}^{XY}}{2} \{ \I \}  + \eps^{m_2} f_{(2)\, Z}^{XY} \{ Z_{2,\eps} \}\right.\\
& \ \ +\left. \eps^{2m_2}\frac{f_{(2)\, A}^{XY}}{2}\{ A_{0,\eps} \} - \eps^{2m_2-1} \frac{f_{(1)\, A}^{XY}}{2} \{ A_{1,\eps} \} \right].
\label{preconXY2}
\end{aligned}\end{equation}
Finally, $X_{1,\eps} \times Y_{2,\eps} \simeq \{ 0 \}$.

In the contraction limit $\eps \to 0$, the OPEs in the subalgebra $\Hc_{G}$ are given by
\begin{equation}
A_{i} \times B_{j} \simeq \begin{cases}
        c_{i+j}\,f^{A}{}^{B}_{\, \I}\{\I\} + f^{A}{}^{B}_{C} \ \{ C_{i+j} \}, \quad &\mathrm{if} \ i+j \leq 1, \\
        0, &\mathrm{if} \ i = j = 1.
        \end{cases}
\end{equation}

For $m_1 \in \left[ 0,\tfrac{1}{2} \right)$, one must have
\begin{equation}
    f_{(1) \, \I}^{AX} = f_{(1)\, B}^{AX} = f_{(1)\, \I}^{XY} = f_{(1)\, A}^{XY} = 0
\end{equation}
for the limit to exist. This condition is equivalent to $\mfg_{1} \times \bar{\mfg}_{1} \subseteq \bar{\mfg}_{1}$.

For $m_1 \in \left[ \tfrac{1}{2},1 \right)$, existence requires
\begin{equation}
    f_{(1)\, \I}^{AX} = f_{(1)\, B}^{AX} = 0.
\end{equation}

The conditions for $m_2 \in \left[ 0,\tfrac{1}{2} \right)$ and $m_2 \in \left[ \tfrac{1}{2},1 \right)$ are the same as those for $m_1$, but with $f_{(1)}$ replaced by $f_{(2)}$. 

In summary, the possible products on the contracted algebra fall into the following cases:
\begin{itemize}
\item[1.] For $m_\ell \notin \{0,\tfrac{1}{2},1 \}$, we have
\begin{equation}\begin{aligned}
A_{0} \times X_{\ell} \simeq  f_{(\ell)\, Y}^{AX} \{ Y_{\ell} \}, \qquad A_{1} \times X_{\ell} \simeq \{ 0 \}, \qquad X_{\ell} \times Y_{\ell} \simeq \{ 0 \}.
\end{aligned}\end{equation}

\item[2.]
For $m_\ell=0$, we have
\begin{equation}\begin{aligned}
A_{0} \times X_{\ell} &\simeq f_{(\ell)\, Y}^{AX} \{ Y_{\ell} \}, \qquad A_{1} \times X_{\ell} \simeq \{ 0 \}, \qquad X_{\ell} \times Y_{\ell} \simeq f^{XY}_{(\ell)\, Z} \{ Z_{\ell} \}.
\end{aligned}\end{equation}

\item[3.] 
For $m_\ell=\frac{1}{2}$, we have 
\begin{equation}
\begin{gathered}
A_{0} \times X_{\ell} \simeq f_{(\ell)\, Y}^{AX} \{ Y_{\ell} \}, \qquad A_{1} \times X_{\ell} \simeq \{ 0 \}, \\
X_{\ell} \times Y_{\ell} \simeq (-1)^{\ell-1}\frac{c_{1} \, f_{(\ell)\, \I}^{XY}}{2} \{ \I \} + (-1)^{\ell-1}\frac{f_{(\ell)\, A}^{XY}}{2} \{ A_{1} \}.
\end{gathered}
\end{equation}

\item[4.]
For $m_\ell = 1$, we have
\begin{equation}
\begin{gathered}
A_{0} \times X_{\ell} \simeq f_{(\ell)\, Y}^{AX} \{ Y_{\ell} \} +(-1)^{\ell-1} \frac{c_{1}\,f_{(\ell)\, \I}^{AX}}{2}\{ \I \} + (-1)^{\ell-1}\frac{f_{(\ell)\, B}^{AX}}{2} \{ B_{\ell} \}, \\ A_{1} \times X_{\ell} \simeq \{ 0 \}, \qquad X_{\ell} \times Y_{\ell} \simeq \{ 0 \}.
\end{gathered}
\end{equation}

\end{itemize}
Following these, the chosen values of the contraction parameters then determine the structure of the resulting Galilean algebra.\qed

The case $m_1=m_2=\frac{1}{2}$ is, in a sense, the natural one. Examples of Lie algebra contractions with these parameter values have been seen before \cite{ORS94,Sfet94,FS94} in the study of WZW models on non-compact Lie groups. It is the only rescaling such that the products between elements in $\bar{G}^{m_{1}}_{1}$ and $\bar{G}^{m_2}_{2}$ only produce fields in $\Hc_{G}$. The resulting product structure is equivalent to that of a $\Z_{2}$-graded Lie algebra, where the even space is given by $\mfh_{G}$, and the odd space is the space $\bar{\mfg}^{\frac{1}{2}}_{1} \oplus \bar{\mfg}^{\frac{1}{2}}_{2}$:
\begin{equation}
    \mfh_{G} \times \mfh_{G} \subseteq \mfh_{G}, \qquad   \mfh_{G} \times \bar{\mfg}^{\frac{1}{2}}_{\ell} \subseteq \bar{\mfg}^{\frac{1}{2}}_{\ell}, \qquad  \bar{\mfg}^{\frac{1}{2}}_{\ell} \times \bar{\mfg}^{\frac{1}{2}}_{\ell} \subseteq \mfh_{G}, \quad \ell = 1,2.
    \label{m1/2}
\end{equation}

We now return to the earlier remark about uniform $\eps$-scaling of the fields in $\bar{G}_\ell$. For our general description, we have only considered rescaling all fields by the same $\eps$-monomial. However, in some cases it may be interesting to rescale generating fields individually. For example, suppose that each field in $\bar{G}_\ell$ is rescaled separately by some $s_{r} \in \R_{\geq 0}$, where $r = 1, \ldots, \vert \bar{G}_\ell \vert $. Then, $X_{1,\eps} \times Y_{1,\eps} = \eps^{s_{1}}X_{(1)} \times \eps^{s_{2}}Y_{(1)}$, so to produce a field $A_{1} \in \Hc_{G}$, we must have $s_{1} +s_{2} = 1$. Thus, there is substantial freedom in choosing $s_{1}$ and $s_{2}$. An example of this is considered in Section \ref{Sec:AsymExamplesSingle}.

We need not restrict ourselves to producing a field $A_{1} \in \mfh_{G}$ in the OPE $X_{1,\eps} \times Y_{1,\eps}$. One can also use individual rescalings to produce interesting graded structures on the fields coming from $\bar{G}_\ell$. Comparing to \eqref{preconXY1}, we see that uniform scaling with $m_{\ell}>0$ cannot produce a field $Z_{\ell}$. However, if the field $Z_{(\ell)}$ is rescaled by $\eps^{s_{1}+s_{2}}$, the term will no longer vanish in the contraction limit. An example where one can introduce such a grading is discussed in Section \ref{Sec:AsymExamplesDouble}.

%***************************************************
\section{Examples with $\Hc \cong \Ac_{2}$}
\label{Sec:AsymExamplesSingle}
%***************************************************
We begin our discussion of examples of asymmetric contractions with the case where one has $\Hc \cong \Ac_{2}$. We use this simplified setting to explore values of $m \neq \frac{1}{2}$. We remark that since $\Hc \cong \Ac_{2}$, the parameter $m_2$ is not required.

%*********************************************
\subsection{The affine Nappi-Witten algebra $\big(\widehat{\mathfrak{gl}}(1) \hookrightarrow \widehat{\mathfrak{sl}}(2)^{\frac{1}{2}} \big)_{G}$}
%*********************************************
Here, we consider the contraction of an affine $\asl(2)$ algebra at level $k_{(1)}$, with an affine $\widehat{\mathfrak{gl}}(1)$ algebra at level $k_{(2)}$, where the asymmetric contraction parameter is $m=\frac{1}{2}$. This example was first discussed in the papers \cite{ORS94,Sfet94}, where the authors considered WZW models on non-compact Lie groups. In particular, they remark that a contraction procedure leads to the affine Nappi-Witten algebra \cite{NW93}, which we denote by $\widehat{H}_{4}$. It has also been realised by alternative contraction procedures, namely in the context of $pp$-waves and Penrose limits \cite{DK03,BDKZ04}. 

We present it again here as a demonstration of a known case of our general construction, and for its physical relevance. In the asymmetric Galilean framework, this contraction (formally) leads to a $2$-parameter generalisation of $\widehat{H}_{4}$. However, we remark that one parameter may freely be set to zero using automorphisms of the algebra (see, e.g.~\cite{DQ08, BKRS20}).

The OPA $\Ac_{1}$, corresponding to the Lie algebra $\asl(2)$ at level $k_{(1)}$, is generated by fields $\{ e,h,f \}$, with nontrivial OPE relations
\begin{equation}
    h \times e \simeq 2 \{ e \}, \quad h \times f \simeq -2 \{ f \}, \quad h \times h \simeq 2k_{(1)} \{ \I \}, \quad e \times f \simeq k_{(1)} \{ \I \} + \{ h \}.
    \label{sl2rels}
\end{equation}

The OPA $\Hc$, corresponding to $\widehat{\mathfrak{gl}}(1)$ at level $k_{(2)}$, is generated by the field $a$, with OPE relation
\begin{equation}\label{gl1rels}
    a \times a \simeq 2k_{(2)} \{ \I \}.
\end{equation}
We see that the field $h \in \asl(2)$ generates such a $\widehat{\mathfrak{gl}}(1)$ subalgebra at level $k_{(1)}$.

The corresponding Galilean algebra $\big(\widehat{\mathfrak{gl}}(1) \hookrightarrow \widehat{\mathfrak{sl}}(2)^{\frac{1}{2}} \big)_{G}$ is generated by the fields $\{ h_{0}, h_{1}, e_{1},f_{1} \}$ with nontrivial OPE relations given 
by
\begin{equation}\begin{gathered}
    h_{0} \times h_{0} \simeq 2k_{0} \{ \I \}, \quad h_{0} \times h_{1} \simeq 2k_{1} \{ \I \}, \\
    h_{0} \times e_{1} \simeq 2 e_{1}, \quad h_{0} \times f_{1} \simeq -2 f_{1}, \quad e_{1} \times f_{1} \simeq \frac{1}{2}k_{1} \{ \I \} + \frac{1}{2} \{ h_{1} \}.
\end{gathered}\end{equation}
It is recognised as the affine Nappi-Witten algebra $\widehat{H}_{4}$ at level $K$ when $k_{0} = 0$ and $k_{1} = K$ \cite{NW93,BKRS20}. 
%***************************************************
\subsection{The asymmetric Galilean Virasoro algebra $\big( \Vir \hookrightarrow (\Vir^{2}_{G})^{n} \big)_{G}$}
\label{AsymVir}
%***************************************************
Our next example is the contraction of the Galilean Virasoro algebra $\Vir^{2}_{G}$ with a Virasoro algebra $\Vir$. This example is well-defined for $m=0$, as the underlying mode algebra of $\Vir^{2}_{G}$ has an abelian ideal. The Galilean Virasoro OPA $\Vir^{2}_{G}$ is generated by the fields $\{ T_{0}, \ T_{1} \}$ and has OPEs given by

\begin{equation}
    T_{i}\times T_{j} \simeq \begin{cases}
        \frac{c}{2}
        \{ \I \} + 2 \{ T_{i+j} \}, \quad &\mathrm{if} \ i+j \leq 1, \\
        0, \quad &\mathrm{otherwise}.
        \end{cases}
\end{equation}

The OPA $\Vir_{G}^{2} \otimes \Vir$ is generated by the fields $ \{ (T_{0})_{(1)}, (T_{1})_{(1)}, T_{(2)} \}$. To perform the contraction (leaving the parameter $m$ free), we form the fields
\begin{equation}
    T_{0,\eps} = (T_{0})_{(1)} + T_{(2)}, \quad T_{1,\eps} =\eps \left((T_{0})_{(1)}- T_{(2)}\right), \quad \bar{T}_{1,\eps} = \eps^{m} (T_{1})_{(1)},
\end{equation}
and central parameters
\begin{equation}
    c_{0,\eps} = (c_{0})_{(1)} + c_{(2)}, \quad c_{1,\eps} = \eps \left( (c_{0})_{(1)} - c_{(2)} \right), \quad \bar{c}_{1,\eps} = \eps^{m} (c_{1})_{(1)}.
\end{equation}

The resulting Galilean algebra $\big( \Vir \hookrightarrow (\Vir^{2}_{G})^{m} \big)_{G}$ is generated by the fields $T_{0}, T_{1}, \bar{T}_{1}$, with nontrivial OPEs
\begin{equation}\begin{aligned}
    T_{i} \times T_{j} &\simeq
    \frac{c_{i+j}}{2}\{ \I \} + 2 \{ T_{i+j} \}, \quad i+j\leq1; \qquad
    T_{0}\times \bar{T}_{1} \simeq \frac{\bar{c}_{1}}{2} \{ \I \} + 2 \{ \bar{T}_{1} \}.
\end{aligned}\end{equation}
For $m=0$, the contraction results in a well-defined algebra with two conformal-weight $2$ quasi-primary fields. This is a novel Galilean structure that cannot be realised using higher-order, or multi-graded, Galilean contractions \cite{RR19,RRR20}.

%*********************************************
\subsection{The asymmetric $N=2$ superconformal algebra $\big(W(1,2) \hookrightarrow \mathrm{SCA}_{2}^{m} \big)_{G}$}
\label{AsymN=2}
%*********************************************
In this example, we consider the $N=2$ superconformal OPA. The algebra is generated by the fields $\{ T, J, G^{+}, G^{-} \}$, where $T, J$ are bosonic, and $G^{\pm}$ are fermionic. The defining nontrivial OPE relations are
\begin{equation}
\begin{gathered}
    T\times T \simeq \frac{c}{2} \{ \I \} + 2 \{ T \} , \quad T \times J \simeq \{ J \}, \quad J \times J \simeq \frac{c}{3} \{ \I \}, \\
    T \times G^{\pm} \simeq \frac{3}{2} \{ G^{\pm} \}, \quad J \times G^{\pm} \simeq \pm \{ G^{\pm} \}, \\
    G^{\pm}\times G^{\mp} \simeq 2\frac{c}{3} \{ \I \} \pm 2 \{J \} + 2 \{ T \}.
\end{gathered}
\end{equation}

The fields $T,J$ generate a bosonic subalgebra which we denote by $W(1,2)$.

The Galilean algebra $\big(W(1,2) \hookrightarrow \mathrm{SCA}_{2}^{m} \big)_{G}$ has elementary fields $\{ T_{0},T_{1},J_{0}, J_{1}, G^{+}_{1}, G^{-}_{1} \}$ with nontrivial OPEs given by
\begin{equation}
\begin{gathered}
    T_{i} \times T_{j} \simeq
    \frac{c_{i+j}}{2} \{ \I \} + 2 \{ T_{i+j} \}, \qquad T_{i} \times J_{j} \simeq \{ J_{i+j} \}, \quad i+j\leq1; \\
    T_{0} \times G^{\pm}_{1} \simeq \frac{3}{2} \{ G^{\pm}_{1} \}, \qquad J_{0} \times G^{\pm}_{1} \simeq \pm \{ G^{\pm}_{1} \},
\end{gathered}
\end{equation}
and $G^{+}_{1} \times G^{-}_{1}$, which is determined by the value of $m$ according to
\begin{equation}
\begin{aligned}\label{fermionOPE}
    G^{+}_{1,\eps} \times G^{-}_{1,\eps} &= \eps^{2m} G^{+}_{(1)} \times G^{-}_{(1)} \\
    &\simeq \eps^{2m} \left( 2\frac{c_{(1)}}{3} \{ \I \} \pm 2 \{ J_{(1)} \} + 2 \{ T_{(1)} \} \right) \\
    &\simeq \eps^{2m}\left(\frac{c_{0}}{3} \{ \I \} \pm \{ J_{0} \} + \{ T_{0} \} \right) + \eps^{2m-1}\left(\frac{c_{1}}{3} \{ \I \} \pm \{ J_{1} \} + \{ T_{1} \} \right).
\end{aligned}
\end{equation}

For $m<\frac{1}{2}$, the contraction is ill-defined. For $m = \frac{1}{2}$, it yields
\begin{equation}
    G^{\pm}_{1}\times G^{\mp}_{1} \simeq \frac{c_{1}}{3} \{ \I \} \pm  \{ J_{1} \} + \{ T_{1} \},
\end{equation}
whereas, for $m > \frac{1}{2}$, $G^{+}_{1}\times G^{-}_{1}  \simeq \{ 0 \}$.

The case of $m=\frac{1}{2}$ has been studied in the literature as the $N = (2,0)$ super-BMS algebra \cite{FMT17}. There has also been interest in $N=1,2,4,$ and $(2,2)$ super-BMS algebras \cite{LM16,BJLMN16,BLN17,CCFR18,BBNN19}.

Asymmetric contractions involving superalgebras provide important examples of contractions where fields in $\bar{G}_{\ell}$ can be individually rescaled, as discussed in Section \ref{Sec:GenFramework}. When rescaled uniformly, the OPE $G^{+}_{1} \times G^{-}_{1}$ is trivial unless $2m=1$, as seen in \eqref{fermionOPE}. However, the fields $G^{\pm}_{(1)}$ may be scaled individually, say by parameters $m_{+},m_{-} \in \R_{\geq0}$. The OPE $G^{+}_{1} \times G^{-}_{1}$ is then trivial unless $m_{+} + m_{-} = 1$. Such contractions have been discussed in the string theory literature \cite{BBCP18,BBCP20}, including the choice $m_{+} = 0$, $m_{-} = 1$. Examples with more fermionic fields, as in the $N=4$ SCAs \cite{N41,N42,STvP88,JR01,JR02}, may allow for more intricate Galilean structures to be introduced using asymmetric contractions.

%*********************************************
\section{Examples with $\Hc \subsetneq \Ac_1$ and $\Hc \subsetneq \Ac_2$}
\label{Sec:AsymExamplesDouble}
%*********************************************
In this section, we present examples where $\Hc_{\ell}$ is a proper subalgebra of $\Ac_\ell$, for each $\ell = 1,2$. We take $m_1=m_2= \frac{1}{2}$ to exhibit the behaviour seen in \eqref{m1/2}, of interest in the literature \cite{ORS94}. The first example involves OPAs associated with affine Lie algebras, which are algebras of Lie-type. The second example involves algebras which are {\em not} Lie-type: the asymmetric contraction of a $W_{4} = W(2,3,4)$ algebra with a $W_{3} = W(2,3)$ algebra. 

%*********************************************
\subsection{$\left( \widehat{\mathfrak{sl}}(3)^{\frac{1}{2}} \hookleftarrow \widehat{\mathfrak{gl}}(1) \hookrightarrow \widehat{\mathfrak{sl}}(2)^{\frac{1}{2}}\right)_{G}$}
%*********************************************
We begin with the contraction of $\Ac_{1} = \widehat{\mathfrak{sl}}(3)$ at level $k_{(1)}$ with $\Ac_{2} = \widehat{\mathfrak{sl}}(2)$ at level $k_{(2)}$, where the equivalent subalgebras are Heisenberg algebras $\Hc_{\ell} = \widehat{\mathfrak{gl}}(1)$. The OPE relations defining the $\asl(2)$ and $\widehat{\mathfrak{gl}}(1)$ OPAs are given in \eqref{sl2rels} and \eqref{gl1rels}, respectively.

The $\asl(3)$ OPA has $8$ elementary fields, namely $\{ h^{1}, h^{2}, e^{r}, f^{r}  \ | \ r = 1, 2, 3 \}$, with nontrivial OPE relations given by
\begin{equation}
\begin{gathered}\label{sl3}
    h^{1} \times e^{1} \simeq 2 \{ e^{1} \}, \qquad h^{1} \times f^{1} \simeq -2 \{ f^{1} \}, \qquad h^{1} \times e^{2} \simeq - \{ e^{2} \}, \qquad h^{1} \times f^{2} \simeq \{ f^{2} \}, \\
    h^{2} \times e^{1} \simeq - \{ e^{1} \}, \qquad h^{2} \times f^{1} \simeq  \{ f^{1} \}, \qquad h^{2} \times e^{2} \simeq  2\{ e^{2} \}, \qquad h^{1} \times f^{2} \simeq -2\{ f^{2} \}, \\
    h^{1} \times e^{3} \simeq \{ e^{3} \}, \qquad h^{1} \times f^{3} \simeq -\{ f^{3} \}, \qquad h^{2} \times e^{3} \simeq \{ e^{3} \}, \qquad h^{2} \times f^{3} \simeq -\{ f^{3} \}, \\
    h^{1} \times h^{1} \simeq 2k_{(1)} \{ \I \}, \qquad h^{2} \times h^{2} \simeq 2k_{(1)} \{ \I \}, \qquad h^{1} \times h^{2} \simeq -k_{(1)} \{ \I \}, \\
    e^{1} \times f^{1} \simeq k_{(1)}\{\I \} + \{ h^{1} \}, \qquad e^{2} \times f^{2} \simeq k_{(1)}\{\I \} + \{ h^{2} \}, \qquad e^{3} \times f^{3} \simeq k_{(1)}\{\I \} + \{ h^{1} \} + \{ h^{2} \}, \\
    e^{1} \times e^{2} \simeq \{ e^{3} \}, \qquad f^{1} \times f^{2} \simeq -\{ f^{3} \}.
\end{gathered}
\end{equation}
The nontrivial OPE relations in the ensuing Galilean algebra are given by
\begin{equation}
\begin{gathered}
    h_{0} \times h_{0} \simeq 2k_{0} \{ \I \}, \quad h_{0} \times h_{1} \simeq 2k_{1} \{ \I \}, \quad h_{\frac{1}{2}} \times h_{\frac{1}{2}} \simeq k_{1} \{ \I \}, \\
    h_{0} \times e^{0}_{2} \simeq 2 \{ e^{0}_{2} \}, \quad h_{0} \times e^{1}_{1} \simeq \{ e^{1}_{1} \}, \quad h_{0} \times e^{2}_{1} \simeq \{ e^{2}_{1} \}, \quad h_{0} \times e^{3}_{1} \simeq 2 \{ e^{3}_{1} \}, \\
    h_{0} \times f^{0}_{2} \simeq -2 \{ f^{0}_{2} \}, \quad h_{0} \times f^{1}_{1} \simeq -\{ f^{1}_{1} \}, \quad h_{0} \times f^{2}_{1} \simeq -\{ f^{2}_{1} \}, \quad h_{0} \times f^{3}_{1} \simeq -2 \{ f^{3}_{1} \}, \\
    e^{0}_{2} \times f^{0}_{2} \simeq -\frac{1}{2} k_{1} \{ \I \} - \frac{1}{2} \{ h_{1} \}, \quad e^{1}_{1} \times f^{1}_{1} \simeq \frac{1}{2} k_{1} \{ \I \} + \frac{1}{4} \{ h_{1} \}, \\
    e^{2}_{1} \times f^{2}_{1} \simeq \frac{1}{2} k_{1} \{ \I \} + \frac{1}{4} \{ h_{1} \}, \quad e^{3}_{1} \times f^{3}_{1} \simeq \frac{1}{2} k_{1} \{ \I \} + \frac{1}{2} \{ h_{1} \}.
\end{gathered}
\end{equation}
In the Galilean algebra, there are no nontrivial relations of the form $e^{r}_{1} \times e^{s}_{1}$ or $f^{r}_{1} \times f^{s}_{1}$ for $r,s \in \{1,2,3 \}$.

%*********************************************
\subsection{$\left( W_{4}^{\frac{1}{2}} \hookleftarrow \Vir \hookrightarrow W_{3}^{\frac{1}{2}} \right)_{G}$}
%*********************************************
Here, we consider an example where the input algebras are not Lie-type. Structure constants of the $W$-algebras $W_{3}$ and $W_{4}$ are algebraic functions of the central parameter $c$, not just constant or linear. Furthermore, OPEs between generating fields produce normally-ordered products of fields.

A method to perform Galilean contractions of $W$-algebras was developed in \cite{GRR14,RR17,RR19}, and many concrete examples of Galilean $W$-algebras have been constructed. 

The $W$-algebra $W_{4} = W(2,3,4)$ is generated by three bosonic fields $\{ T,W,U \}$, where $T$ generates a Virasoro subalgebra, and $W$ and $U$ are primary fields of conformal weight $3$ and $4$, respectively. The nontrivial OPEs between generating fields are
\begin{equation}
\begin{gathered}
T \times T \simeq \frac{c}{2} \{ \I \} + 2 \{ T \}, \quad T \times W \simeq 3 \{ W \}, \quad T \times U \simeq 4 \{ U \}, \\
W \times W \simeq \frac{c}{3} \{ \I \} + 2 \{ T \} + \frac{32}{5c+22} \{ \Lambda^{2,2} \} + \frac{4}{3}\lambda \{ U \}, \\
W \times U \simeq \lambda \{ W \} + \frac{52}{7c+114} \{ \Lambda^{2,3} \} + \frac{1}{c+2}\{ \Lambda^{2',3} \}, \\
U \times U \simeq \frac{c}{4} \{ \I \} + 2 \{ T \} + \frac{42}{5c+22}\{ \Lambda^{2,2} \} - \frac{3\lambda(c^2+c+218)}{(c+2)(7c+114)} \{ U \} \\
+ \frac{45(5c+22)}{2(c+2)(7c+114)}\{ \Lambda^{3,3} \} - \frac{4\lambda}{c+2} \{ \Lambda^{3,3} \} + \frac{96(9c-2)}{(c+2)(5c+22)(7c+114)} \{ \Lambda^{2,2,2} \} \\
+ \frac{3(19c^2 - 844c - 2484)}{4(c+2)(5c+22)(7c+114)} \{ \Lambda^{2'',2} \},
\end{gathered}
\end{equation}
where
\begin{equation}\label{lambda}
\lambda = \sqrt{\frac{3(c+2)(7c+114)}{(c+7)(5c+22)}}.
\end{equation}

Here, $\Lambda^{r,s,\ldots,t}$ denotes a quasi-primary field associated with the normally-ordered product of generating fields with conformal weights $r,s,\ldots,t$, respectively. Our convention is that the normally-ordered product of more than two fields is right-nested. Primes accompanying an index denote the number of derivatives acting on that component field. For example,
\begin{equation}
\Lambda^{2'',2} = \left( \partial^2 T T \right) - \frac{5}{18}\partial^2 \Lambda^{2,2} - \frac{1}{42}\partial^4 T.
\end{equation}
We choose to normalise so that the term corresponding to the superscript has coefficient $1$. Given a normally-ordered product of generating fields and their derivatives, the corresponding normally-ordered field is determined, up to normalisation, by its conformal transformation properties. 

Similarly, the $W$-algebra $W_{3}$ is generated by fields $\{ T, W  \}$, where $T$ generates a Virasoro subalgebra, and $W$ is a primary field of conformal weight $3$. The OPE relations for $W_{3}$ are given by
\begin{equation}
\begin{gathered}
T \times T \simeq \frac{c}{2} \{ \I \} + 2 \{ T \}, \quad T \times W \simeq 3 \{ W \}, \\
W \times W \simeq \frac{c}{3} \{ \I \} + 2 \{ T \} + \frac{32}{5c+22} \{ \Lambda^{2,2} \}.
\end{gathered}
\end{equation}

We remark that the $W_{3}$ algebra is not a subalgebra of the $W_{4}$ algebra.

The change of basis for the generating fields in this example is given by
\begin{equation}
T_{i,\eps} = \eps^{i} \left( T_{(1)} + (-1)^i T_{(2)} \right), \quad W_{\ell,\eps} = \eps^{\frac{1}{2}} W_{(\ell)}, \quad U_{1,\eps} = \eps^{\frac{1}{2}} U_{(1)}, \quad i = 0,1, \ \ell = 1,2.
\end{equation}

The asymmetrically contracted algebra is then generated by the fields $\{ T_{0}, T_{1}, W_{1}, W_{2}, U_{1} \}$, and the Lie-type OPE relations are given by
\begin{equation}
\begin{gathered}
    T_{i} \times T_{j} \simeq \frac{c_{i+j}}{2} \{ \I \} + 2 \{ T_{i+j} \}, \quad \mathrm{if} \ i+j \leq 1, \\
    T_{0}\times W_{\ell} \simeq 3 \{ W_{\ell} \}, \qquad T_{0}\times U_{1} \simeq  4 \{ U_{1} \}, \\
    T_{1} \times T_{1} \simeq T_{1} \times W_{\ell} \simeq T_{1} \times U_{1} \simeq \{ 0 \}.
\end{gathered}\end{equation}
The Galilean Virasoro subalgebra is generated by $T_{0}, T_{1}$, and the fields $W_{\ell}, U_{1}$ are primary with respect to the Virasoro field $T_{0}$. The remaining OPEs are determined by applying the techniques detailed in \cite{RR17,RR19}.

To illustrate, consider the OPE 
\begin{equation}\begin{aligned}\label{W1W1}
    W_{1,\eps} \times W_{1,\eps} &= \eps^{1} \left( W_{(1)} \times W_{(1)} \right) \\
    &\simeq \eps \left( \frac{c_{(1)}}{3} \{ \I \} + 2 \{ T_{(1)} \} + \frac{32}{5c_{(1)}+22} \{ \Lambda^{2,2}_{(1)} \} +\frac{4\lambda_{(1)}}{3} \{ U_{(1)} \} \right).
\end{aligned}\end{equation}
Applying \eqref{asyminvbasis} to $c_{(1)}$ and expanding as a power series about $\eps$ small, we have
\begin{equation}
    \frac{32}{22+5c_{(1)}}
    =\frac{64}{5c_{1}}\eps - \frac{64(44+5c_{0})}{25c_{1}^{2}} \eps^2 + \mathcal{O}(\eps^{3}),
\end{equation}
and
\begin{equation}
    \frac{4\lambda_{(1)}}{3}=4\sqrt{\frac{7}{15}} + \frac{964}{5\sqrt{105}c_{1}}\eps + \frac{2(16870c_{0}+434037)}{175\sqrt{105}c_{1}^{2}}\eps^2 + \mathcal{O}(\eps^{3}),
\end{equation}
where $\lambda_{(1)}$ is the expression \eqref{lambda} as a function of $c_{(1)}$. The fields appearing on the right-hand side of \eqref{W1W1} are expanded using the inverse maps \eqref{asyminvbasis}.

The resulting expressions are combined, and the limit $\eps \to 0$ is taken, resulting in
\begin{equation}
    W_{1} \times W_{1} \simeq \frac{c_{1}}{6} \{ \I \} + \{ T_{1} \} + \frac{16}{5c_{1}} \{ \Lambda^{2,2}_{1,1} \},
\end{equation}
where the subscripts on the field $\Lambda^{2,2}_{1,1}$ denote the corresponding subscripts on the leading normally-ordered term, that is, $\Lambda^{2,2}_{1,1}= \left( T_{1} T_{1} \right)$. The remaining nontrivial products on the contracted algebra are given by
\begin{equation}
\begin{gathered}
W_{2} \times W_{2} \simeq -\frac{c_{1}}{6} \{ \I \} - \{ T_{1} \} - \frac{16}{5c_{1}} \{ \Lambda^{2,2}_{1,1} \}, \\
    U_{1} \times U_{1} \simeq \frac{c_{1}}{8} \{ \I \} + \{ T_{1} \} + \frac{21}{5c_{1}} \{ \Lambda^{2,2}_{1,1} \} + \frac{3456}{280(c_{1})^2} \{ \Lambda^{2,2,2}_{1,1,1} \} + \frac{57}{280c_{1}} \{ \Lambda^{2'',2}_{1,1} \}.
\end{gathered}
\end{equation}

In this case, OPEs between primary fields produce only Galilean Virasoro fields and normally-ordered products thereof. We also could have chosen to rescale $W$ and $U$ separately in the $W_{4}$ algebra. For example, we could scale $W_{(1)}$ by $\eps^{\frac{1}{2}}$ and $U_{(1)}$ by $\eps^{1}$, thereby introducing a Galilean-type structure amongst fields coming from $\bar{G}_{1}$. With these changes in rescaling, the non-Lie-type OPEs become 
\begin{equation}
    W_{1}\times W_{1} \simeq   \frac{c_{1}}{6} \{ \I \} + \{ T_{1} \} + \frac{16}{5c_{1}} \{ \Lambda^{2,2}_{1,1} \} + 4\sqrt{\frac{7}{15}}\{ U_{1} \}, \qquad W_{1} \times U_{1} \simeq U_{1} \times U_{1} \simeq \{ 0 \}.
\end{equation}

%*********************************************
\section{The asymmetric Sugawara construction}
\label{Sec:AsymSuga}
%*********************************************
Here, we want to understand when it is possible to construct a Virasoro field from an asymmetric Galilean affine Lie algebra in a process analogous with the Sugawara construction. As Galilean contracted algebras contain an abelian ideal, they are not semisimple. There is a substantial amount of literature devoted to developing a Sugawara construction for non-semisimple Lie algebras \cite{HK89,Moham94,Sfet94,HKOC96,FS94}. 

The goal of the construction is to build a field from bilinears and derivatives of the currents which satisfies the Virasoro algebra OPE relations, and with respect to which the currents are primary fields of conformal weight $1$. Such a field naturally satisfies all the constraints placed by conformal symmetry on the stress-energy tensor of a particular model. However, we cannot say exactly for what model the field would be a stress-energy tensor, as our interest is purely algebraic. It would be of interest in the physics literature to identify and understand the models for which the constructed field is in fact the stress-energy tensor.

It was shown in \cite{Moham94, FS94} that a necessary and sufficient condition for such a field to exist for an infinite-dimensional Lie algebra is that the algebra possesses a non-degenerate invariant symmetric bilinear form, here denoted by $\Omega=(\Omega^{ab})$. 

From \cite{FS94}, we have that the general form of $\Omega$ is given by 
\begin{equation}
    \Omega= 2g+ \kappa,
\end{equation}
where $\kappa$ is the Killing form ($\kappa^{ab} = f^{a}{}^{c}_{d}f^{b}{}^{d}_{c}$), and $g=(g^{ab})$ is an invariant bilinear form arising in the OPEs of the current algebra:
\begin{equation}
    J^a(z) J^b(w) \sim \frac{g^{ab}}{(z-w)^2} + f^{a}{}^{b}_{c} \frac{J^{c}(w)}{z-w}.
\end{equation}
The invariance of $g$ is necessary for the associativity of the OPE. For simple algebras, $\Omega$ is a scalar multiple of the Killing form:
\begin{equation}
    \Omega= 2(k+h^{\vee}) \kappa,
\end{equation}
and the corresponding Sugawara operator $T(z)$ is then given by the well-known expression
\begin{equation}
    T(z) = \sum_{a,b}\Omega_{ab}\left( J^aJ^b\right)(z) = \frac{1}{2(k + h^{\vee})}\sum_{a,b}\kappa_{ab}\left( J^aJ^b\right)(z),
\end{equation}
where $(\Omega_{ab})$ and $(\kappa_{ab})$ are the inverses of $(\Omega^{ab})$ and $(\kappa^{ab})$, respectively.

We begin with affine Lie OPAs $\widehat{\mfg}_1, \ \widehat{\mfg}_2$, each with an embedded affine Lie subalgebra $\mfhh_{\ell}$. Our notation for the currents of the algebra $\widehat{\mfg}_\ell$ is as follows: Currents in $\widehat{\mfh}_\ell$ are labelled with Roman letter group indices and are denoted by $J^{a}_{(\ell)}, \ a \in 1,\ldots, \dim\mfh$; the remaining currents are labelled with Greek letter group indices and are denoted by $J^{\alpha}_{(\ell)}, \ \alpha = 1, \ldots, \dim\mfg_{\ell}- \dim\mfh$; where $\mfg_{\ell}$ and $\mfh$ are the Lie algebras underlying $\widehat{\mfg}_{\ell}$ and $\widehat{\mfh}$, respectively.

The algebra $\widehat{\mfg}_{\ell}$ then has OPEs given in their most general form by 
\begin{equation}\begin{gathered}
    J^{a}_{(\ell)} \times J^{b}_{(\ell)} \simeq M^{ab} \,k_{(\ell)} \{ \I \} + f^{a}{}^{b}_{c} \{ J^{c}_{(\ell)} \}, \\
    J^{a}_{(\ell)} \times J^{\alpha}_{(\ell)}  \simeq M^{a\alpha}_{(\ell)} \, k_{(\ell)} \{ \I \} + f^{a\alpha}_{(\ell)\, b} \{ J^{b}_{(\ell)} \} + f^{a\alpha}_{(\ell)\, \beta} \{ J^{\beta}_{(\ell)} \},  \\
    J^{\alpha}_{(\ell)} \times J^{\beta}_{(\ell)}  \simeq M^{\alpha \beta}_{(\ell)} \, k_{(\ell)} \{ \I \} + f^{\alpha \beta}_{(\ell)\, a} \{ J^{a}_{(\ell)} \} + f^{\alpha \beta}_{(\ell)\, \gamma} \{ J^{\gamma}_{(\ell)} \}.
\end{gathered}\end{equation}
As before, we have dropped the subscript label for the structure constants of the subalgebras $\mfhh_{\ell}$, namely $M^{ab}$ and $f^{a}{}^{b}_{c}$.

We assume that each of the input algebras admits a Sugawara construction. We want to determine conditions under which the asymmetric Galilean algebra $\left( \widehat{\mfg}_1^{\, m_1} \hookleftarrow  \widehat{\mfh} \hookrightarrow \widehat{\mfg}_2^{\, m_2}\right)_{G}$ admits a non-degenerate symmetric invariant bilinear form $\Omega$. 

Before taking the limit $\eps \to 0$, we have the following $\eps$-dependent OPEs:
\begin{equation}
    J^{a}_{i,\eps} \times J^{b}_{j,\eps} \simeq M^{ab}k_{i+j, \eps} \{ \I \} + f^{ab}{}_{c} \{ J^{c}_{i+j,\eps} \},
\end{equation}
\begin{equation}\begin{aligned}
    J^{a}_{i,\eps} \times J^{\alpha}_{1,\eps} &\simeq \frac{M^{a \alpha}_{(1)}}{2}\left( \eps^{i+m_1}k_{0,\eps} + \eps^{i+m_1-1}k_{1,\eps} \right) \{ \I \} + f^{a\alpha}_{(1)\, \beta} \eps^{i} \{ J^{\beta}_{1,\eps} \}  \\
    & \ + \frac{f^{a \alpha}_{(1)\, b}}{2} \left( \eps^{i+m_1} \{ J^{b}_{0,\eps} \} + \eps^{i+m_1-1} \{ J^{b}_{1,\eps} \} \right),
\end{aligned}\end{equation}
\begin{equation}\begin{aligned}
    J^{a}_{i,\eps} \times J^{\alpha}_{2,\eps} &\simeq (-1)^{i} \left[ \frac{M^{a \alpha}_{(2)}}{2}\left( \eps^{i+m_2}k_{0,\eps} -\eps^{i+m_2-1}k_{1,\eps}  \right) \{ \I \} + f^{a\alpha}_{(2)\, \beta} \eps^{i} \{ J^{\beta}_{2,\eps} \} \right. \\
    & \ + \left. \frac{f^{a \alpha}_{(2)\, \beta}}{2} \left( \eps^{i+m_2} \{ J^{b}_{0,\eps} \} - \eps^{i+m_2-1} \{ J^{b}_{1,\eps} \}  \right) \right],
\end{aligned}\end{equation}
\begin{equation}\begin{aligned}
    J^{\alpha}_{1,\eps} \times J^{\beta}_{1,\eps} &\simeq \frac{M^{\alpha \beta}_{(1)}}{2}\left( \eps^{2m_1}k_{0,\eps} + \eps^{2m_1-1}k_{1,\eps} \right) \{ \I \} + f^{\alpha \beta}_{(1)\, \gamma} \eps^{m_1} \{ J^{\gamma}_{1,\eps} \}  \\
    & \ + \frac{f^{\alpha \beta}_{(1)\, a}}{2} \left( \eps^{2m_1} \{ J^{a}_{0,\eps} \} + \eps^{2m_1-1} \{ J^{a}_{1,\eps} \} \right),
\end{aligned}\end{equation}
and
\begin{equation}\begin{aligned}
    J^{\alpha}_{2,\eps} \times J^{\beta}_{2,\eps} &\simeq \frac{M^{\alpha \beta}_{(2)}}{2}\left( \eps^{2m_2}k_{0,\eps} -\eps^{2m_2-1}k_{1,\eps} \right) \{ \I \} + f^{\alpha \beta}_{(1)\, \gamma} \eps^{m_2} \{ J^{\gamma}_{2,\eps} \} \\
    & \ + \frac{f^{\alpha \beta}_{(1)\, a}}{2} \left( \eps^{2m_2} \{ J^{a}_{0,\eps} \} - \eps^{2m_2-1} \{ J^{a}_{1,\eps}\} \right).
\end{aligned}\end{equation}

The form $g$ can be read off these relations. Relative to the ordered basis $\{ J^{a}_{0,\eps}, J^{a}_{1,\eps}, J^{\alpha}_{1,\eps}, J^{\alpha}_{2,\eps} \}$, $(g^{\bullet \bullet})$ is a block matrix given by
\begin{equation}
    (g^{\bullet \bullet}) = 
    \begin{pmatrix}
    k_{0,\eps}M^{ab} & k_{1,\eps}M^{ab} & \boxed{1} & \boxed{3} \\
    k_{1,\eps}M^{ab} & 0 & \boxed{2} & \boxed{4} \\
    \boxed{1} & \boxed{2} & \boxed{5} & 0 \\
    \boxed{3} & \boxed{4} & 0 & \boxed{6} \\
    \end{pmatrix},
\end{equation}
where 
\begin{equation}\begin{aligned}
    \boxed{1} &= \frac{M^{a\alpha}_{(1)}}{2}\left( k_{0,\eps} \eps^{m_1} + k_{1,\eps}\eps^{m_1-1} \right), & \boxed{2} &=  \frac{M^{a\alpha}_{(1)}}{2}\left( k_{0,\eps} \eps^{m_1+1} + k_{1,\eps}\eps^{m_1} \right), \\
    \boxed{3} &=  \frac{M^{a\alpha}_{(2)}}{2}\left( k_{0,\eps} \eps^{m_2} - k_{1,\eps}\eps^{m_2-1} \right), & \boxed{4} &=  \frac{M^{a\alpha}_{(2)}}{2}\left( -k_{0,\eps} \eps^{m_2+1} + k_{1,\eps}\eps^{m_2} \right), \\
    \boxed{5} &=  \frac{M^{\alpha \beta}_{(1)}}{2}\left( k_{0,\eps} \eps^{2m_1} + k_{1,\eps}\eps^{2m_1-1} \right), & \boxed{6} &=  \frac{M^{\alpha \beta}_{(2)}}{2}\left( k_{0,\eps} \eps^{2m_2} - k_{1,\eps}\eps^{2m_2-1} \right).
\end{aligned}\end{equation}
If the contraction parameters $m_1$ and $m_2$ are such that the limit $\eps \to 0$ does not exist for general $M_{(\ell)}^{\bullet \bullet}$ in the expressions above, then we are forced only  to consider algebras for which some blocks in $M_{(\ell)}^{\bullet \bullet}$ (labelled by the upper indices) are zero, allowing the limit to be taken:
\begin{itemize}
    \item For $m_1 = 0$, we require $M^{a \beta}_{(1)}=M^{\alpha \beta}_{(1)}=0$, in which case $\boxed{1} = \boxed{2} = \boxed{5} = 0$.
   \item For $m_1 = \frac{1}{2}$, we require $M^{a \beta}_{(1)}=0$, in which case $\boxed{1} = \boxed{2} = 0$.
    \item For $m_1 = 1$ we have $\boxed{2} = \boxed{5} = 0$ in the contraction limit.
    \item For $m_2 = 0$, we require $M^{a \beta}_{(2)}=M^{\alpha \beta}_{(2)}=0$, in which case $\boxed{3} = \boxed{4}=\boxed{6}=0$.
    \item For $m_2 = \frac{1}{2}$, we require $M^{a\beta}_{(2)}=0$, in which case $\boxed{3} = \boxed{4}=0$.
    \item For $m_2 = 1$, we have $\boxed{4} = \boxed{6}=0$ in the contraction limit.
\end{itemize}

The Killing form on the pre-contraction algebra with respect to the same ordered basis is given by the block matrix
\begin{equation}
    (\kappa^{\bullet \bullet}) = 
    \begin{pmatrix}
     \kappa^{a_{0}a_{0}}& \kappa^{a_{0}a_{1}} & \kappa^{a_{0}\alpha_{1}} & \kappa^{a_{0}\alpha_{2}} \\
     \kappa^{a_{0}a_{1}} & \kappa^{a_{1}a_{1}} & \kappa^{a_{1}\alpha_{1}}  &  \kappa^{a_{1}\alpha_{2}} \\
    \kappa^{\alpha_{1}a_{0}} & \kappa^{\alpha_{1}a_{1}} & \kappa^{\alpha_{1}\alpha_{1}} & \kappa^{\alpha_{1}\alpha_{2}}  \\
    \kappa^{\alpha_{2}a_{0}} & \kappa^{\alpha_{2}a_{1}} & \kappa^{\alpha_{2}\alpha_{1}} & \kappa^{\alpha_{2}\alpha_{2}}  \\
    \end{pmatrix},
\end{equation}
where we have introduced the shorthand $a_{i}$ for currents $J^{a}_{i}$, and $\alpha_{1\ell}$ for $J^{\alpha}_{\ell}$. The entries are given by
\begin{equation}
    \begin{aligned}
       \kappa^{a_{0}a_{0}} &= 2(f^{a}{}^{b}_{c})^2 + (f_{(1)\, \beta}^{a\alpha})^2 + (f_{(2)\, \beta}^{a\alpha})^2, & \kappa^{a_{0}\alpha_{1}} &= -\eps^{m_1} \left((f^{a}{}^{b}_{c}f_{(1)\, b}^{a \alpha} + 
   f_{(1)\, \beta}^{a \alpha} (f_{(1)\, b}^{a \alpha} - 
      f_{(1)\, \gamma}^{\alpha \beta})\right), \\
    \kappa^{a_{1}a_{1}} &= \eps^2 \left((f_{(1)\, \beta}^{a \alpha})^2 + f_{(2)\, \beta}^{a \alpha})^2\right), & \kappa^{a_{1}\alpha_{1}} &= -\frac{1}{2} \eps^{m_1+1} \left( f^{a}{}^{b}_{c} f_{(1)\, b}^{a \alpha} + 
   2f_{(1)\, \beta}^{a \alpha} (f_{(1)\, b}^{a \alpha} - 
      f_{(1)\, \gamma}^{\alpha \beta}) \right), \\
    \kappa^{a_{0}a_{1}} &= \eps \left((f_{(1)\, \beta}^{a\alpha})^2 - f_{(2)\, \beta}^{a\alpha})^2\right), & \kappa^{a_{0}\alpha_{2}} &= -\eps^{m_2} \left( f^{a}{}^{b}_{c} f_{(2)\, b}^{a \alpha} + 
   f_{(2)\, \beta}^{a \alpha} (f_{(2)\, b}^{a \alpha} - 
      f_{(2)\, \gamma}^{\alpha \beta}) \right),\\
   \kappa^{\alpha_{1}\alpha_{2}} &= 0, & \kappa^{a_{1}\alpha_{2}} &= \frac{1}{2} \eps^{m_2+1} \left(f^{a}{}^{b}_{c} f_{(2)\, b}^{a \alpha} + 
   2f_{(2)\, \beta}^{a \alpha} (f_{(2)\, b}^{a \alpha} - 
      f_{(2)\, \gamma}^{\alpha \beta}) \right), \\
   \kappa^{\alpha_{2}\alpha_{2}} &= \eps^{2m_2} \left((f_{(2)\, b}^{\alpha a})^2 +
   f_{(2)\, \beta}^{\alpha a} f_{(2)\, a}^{\alpha \beta} + (f_{(2)\, \gamma}^{\alpha\beta})^2\right), & \kappa^{\alpha_{1}\alpha_{1}} &= \eps^{2m_1} \left( (f_{(1)\, b}^{a \alpha})^2 - 
   2 f_{(1)\, \beta}^{a \alpha} f_{(1)\, a}^{\alpha \beta} + (f_{(1)\, \gamma}^{\alpha \beta})^2\right); \nn
    \end{aligned}
\end{equation}
with the remaining ones following from the symmetry of $(\kappa^{\bullet \bullet})$.

The entries $\kappa^{a_{0}a_{1}}, \ \kappa^{a_{1}a_{1}}, \  \kappa^{a_{1}\alpha_{1}}, \ \kappa^{a_{1}\alpha_{2}},$ all vanish in the contraction limit. We remark that it may occur that, for the contraction limit to exist for particular values of the contraction parameters, some structure constants in the above expression must be zero.

There are only four possible structures for the Killing form in the contraction limit. If $m_1 = m_2 = 0$, we have
\begin{equation}
    (\kappa^{\bullet \bullet}) = 
    \begin{pmatrix}
     2 (f^{a}{}^{b}_{c})^2 + (f_{(1)\, \beta}^{a \alpha})^2 + 
 (f_{(2)\, \beta}^{a \alpha})^2& 0 & f_{(1)\, \beta}^{a \alpha} f_{(1)\, \gamma}^{\alpha \beta} & f_{(2)\, \beta}^{a \alpha} f_{(2)\, \gamma}^{\alpha \beta} \\
     0 & 0 & 0 &  0 \\
    f_{(1)\, \beta}^{a \alpha} f_{(1)\, \gamma}^{\alpha \beta} & 0 & (f_{(1)\, \gamma}^{\alpha \beta})^2 & 0  \\
    f_{(2)\, \beta}^{a \alpha} f_{(2)\, \gamma}^{\alpha \beta} & 0 & 0 & (f_{(2)\, \gamma}^{\alpha \beta})^2 \\
    \end{pmatrix}.
\end{equation}
If $m_1 = 0$ and $m_2 \neq 0$, we have
\begin{equation}
    (\kappa^{\bullet \bullet}) = 
    \begin{pmatrix}
     2 (f^{a}{}^{b}_{c})^2 + (f_{(1)\, \beta}^{a \alpha})^2 + 
 (f_{(2)\, \beta}^{a \alpha})^2& 0 & f_{(1)\, \beta}^{a \alpha} f_{(1)\, \gamma}^{\alpha \beta} & 0 \\
     0 & 0 & 0 &  0 \\
    f_{(1)\, \beta}^{a \alpha} f_{(1)\, \gamma}^{\alpha \beta} & 0 & (f_{(1)\, \gamma}^{\alpha \beta})^2 & 0  \\
    0 & 0 & 0 & 0 \\
    \end{pmatrix}.
\end{equation}
Similarly, if $m_2 = 0$ and $m_1 \neq 0$, we have
\begin{equation}
    (\kappa^{\bullet \bullet}) = 
    \begin{pmatrix}
     2 (f^{a}{}^{b}_{c})^2 + (f_{(1)\, \beta}^{a \alpha})^2 + 
 (f_{(2)\, \beta}^{a \alpha})^2& 0 & 0 & f_{(2)\, \beta}^{a \alpha} f_{(2)\, \gamma}^{\alpha \beta} \\
     0 & 0 & 0 &  0 \\
    0 & 0 & 0 & 0  \\
    f_{(2)\, \beta}^{a \alpha} f_{(2)\, \gamma}^{\alpha \beta} & 0 & 0 & (f_{(2)\, \gamma}^{\alpha \beta})^2 \\
    \end{pmatrix}.
\end{equation}
Finally, if $m_1 \neq 0$ and $m_2 \neq 0$, the Killing form becomes
\begin{equation}
    (\kappa^{\bullet \bullet}) = 
    \begin{pmatrix}
     2 (f^{a}{}^{b}_{c})^2 + (f_{(1)\,\beta}^{a \alpha})^2 + (f_{(2)\, \beta}^{a \alpha})^2& 0 & 0 & 0 \\
     0 & 0 & 0 &  0 \\
   0 & 0 & 0 & 0  \\
    0 & 0 & 0 & 0 \\
    \end{pmatrix}.
\end{equation}
From this, we see that the form $\Omega= 2g+ \kappa$ can only be invertible if $m_1,m_2 \in \{ 0, \tfrac{1}{2}, 1 \}$ and hence, only in these cases is a Sugawara construction possible on the contracted algebra, subject to the exact values of the structure constants $M$ and $f$. By the results of \cite{FS94}, the central charge of the resulting Virasoro algebra generated by the Sugawara field is given by
\begin{equation}
    c = \dim\mfg_{1}+ \dim\mfg_{2}- \Omega_{rs}\kappa^{rs}.
\end{equation}

In the situation that a Sugawara operator cannot be constructed (that is, one cannot find a nondegenerate invariant bilinear form), one could still consider an extension of the field algebra whereby one introduces an additional Virasoro field. This would result in the extended algebra being a conformal algebra; however, the resulting Virasoro field cannot be realised in terms of bilinears and derivatives of the currents which make up the original affine Galilean algebra.

We remark that in the papers \cite{NW93,ORS94,Sfet94,FS94}, the authors consider algebras which can be realised using the general description above. In particular, examples given in \cite{NW93,ORS94,Sfet94} correspond to cases with $\widehat{\mfg}_{2} \cong \widehat{\mfh}_{2}$ and $m = \frac{1}{2}$.

\section{Discussion}
In this paper, we have introduced a framework for performing Galilean-type contractions where the input algebras are not alike, relaxing a prior restriction on such contractions. Although our presentation focuses on the case of two input algebras, the generalisation to any number of input algebras is straightforward. Such algebras have higher-order Galilean subalgebras \cite{RR19} with OPEs between the fields in $\overline{G}^{m_\ell}_{\ell}$ determined by \Cref{mainprop}.

Our asymmetric Galilean contraction procedure provides a unifying framework for several interesting examples in the CFT literature, from early work on non-compact WZW models to current work on tensionless strings and super-BMS symmetries. In a sense, Galilean contraction procedures provide an opposite construction to semigroup expansions which also give rise to Galilean subalgebras \cite{IRS06,IPRS09,CCRS18}. That is, the asymmetric contraction constrains a larger symmetry rather than enlarging a smaller symmetry. However, the question of whether every semigroup expanded algebra can be obtained by a suitable choice of contraction remains open.

In Section \ref{Sec:AsymSuga}, we have demonstrated that for particular values of the contraction parameters, namely $m_1,m_2\in \{ 0, \tfrac{1}{2},1 \}$, asymmetric Galilean algebras arising from affine Lie algebras admit a Sugawara construction. 
It remains to be determined whether one can also construct an accompanying field of conformal weight $2$ which, along with the Sugawara field, generates a Galilean Virasoro algebra $\Vir^{2}_{G}$ (see Section \ref{Sec:AsymExamplesSingle}). Such a field arises in the Sugawara construction for other Galilean algebras (see \cite{RR17,RR19,RRR20}); however, those results rely upon concrete expressions for the Sugawara fields of the input algebras.

We have not considered the representation theory of asymmetric Galilean algebras in this paper. However, representations of algebras that arise in the asymmetric Galilean framework have been studied elsewhere. For example, the representation theory of the affine Nappi-Witten algebra $\widehat{H}_{4}$ is considered in \cite{KK93,JR04,BJP11,BKRS20}. In particular, the authors of \cite{BKRS20} demonstrate the existence of indecomposable yet reducible representations, indicating that $\widehat{H}_{4}$ may encode the symmetries of a logarithmic CFT.

It would be interesting to see if logarithmic behaviour can arise from the action of the Galilean Virasoro algebra, as the grading on Galilean algebras produces a non-diagonalisable action of the generators. For example, it is well known that the algebra $\widehat{H}_{4}$ admits a Sugawara construction. In addition, one can also construct a Galilean partner field such that these fields generate an action of the Galilean Virasoro algebra $\Vir^2_{G}$.

\section*{Acknowledgements}
CR was supported by Australian Research Council grants DP170103265 and DP200100067, and a University of Queensland Research Award.
JR was supported by the Australian Research Council under the Discovery Project scheme, project numbers DP160101376 and DP200102316.
Part of this work was done while ER was visiting the University of Queensland in September 2019, 
with support from the SMRI International Visitor program at the University of Sydney and the University of Queensland.

\printbibliography

@Article{N42,
  author      = {Ademollo, M. and Brink, L. and D’Adda, A. and D’Auria, R. and Napolitano, E. and Sciuto, S. and Del Giudice, E. and Di Vicchia, P. and Ferrara, S. and Gliozzi, F. and Musto, R. and Pettorino, R.},
  journal     = {Nucl. Phys. B},
  title       = {Dual string models with non-abelian colour and flavour symmetries},
  year        = {1976},
  pages       = {297–-316},
  volume      = {114},
}

@Article{N41,
  author      = {Ademollo, M. and Brink, L. and D’Adda, A. and D’Auria, R. and Napolitano, E. and Sciuto, S. and Del Giudice, E. and Di Vicchia, P. and Ferrara, S. and Gliozzi, F. and Musto, R. and Pettorino, R.},
  journal     = {Phys. Lett. B},
  title       = {Supersymmetric strings and colour confinement},
  year        = {1976},
  pages       = {105-–110},
  volume      = {62},
}

@Article{JR01,
  author      = {Rasmussen, J.},
  journal     = {Nucl. Phys. B},
  title       = {Comments on $N = 4$ superconformal algebras},
  year        = {2001},
  pages       = {634-–650},
  volume      = {593},
  eprint      = {hep-th/0003035},
  eprinttype  = {arXiv},
}

@Article{JR02,
  author      = {Rasmussen, J.},
  journal     = {J. Phys. A},
  title       = {A non-reductive $N = 4$ superconformal algebra},
  year        = {2002},
  pages       = {2037-–2044},
  volume      = {35},
  eprint      = {hep-th/0108054},
  eprinttype  = {arXiv},
}

@Article{STvP88,
  author      = {Sevrin, A. and Troost, W. and Van Proyen, A.},
  journal     = {Phys. Lett. B},
  title       = {Superconformal algebras in two dimensions with $N=4$},
  year        = {1988},
  pages       = {447--450},
  volume      = {208},
}

@Article{JR04,
  author      = {Rasmussen, J.},
  journal     = {Afr. J. Math. Phys.},
  title       = {On string backgrounds and (logarithmic) CFT},
  year        = {2004},
  pages       = {171--175},
  volume      = {1},
  eprint      = {hep-th/0404226},
  eprinttype  = {arXiv},
}

@Article{IPRS09,
  author      = {Izuarieta, F. and P{\'e}rez, A. and Rodr{\'i}guez, E. and Salgado, P.},
  journal     = {J. Math. Phys.},
  title       = {Dual formulation of the Lie algebra $S$-expansion procedure},
  year        = {2009},
  pages       = {073511},
  volume      = {50},
  eprint      = {0903.4712},
  eprintclass = {hep-th},
  eprinttype  = {arXiv},
}

@Article{IRS06,
  author      = {Izaurieta, F. and Rodr{\'i}guez, E. and Salgado, P.},
  journal     = {J. Math. Phys.},
  title       = {Expanding Lie (super)algebras through Abelian semigroups},
  year        = {2006},
  pages       = {123512},
  volume      = {47},
  eprint      = {hep-th/0606215},
  eprinttype = {arXiv},
}

@Article{BBNN19,
  author      = {Banerjee, N. and Bhattacharjee, A. and Neetu and Neogi, T.},
  journal     = {JHEP},
  title       = {New $N=2$ SuperBMS${}_3$ algebra and invariant dual theory for 3D supergravity},
  year        = {2019},
  pages       = {122},
  volume      = {11},
  eprint      = {1905.10239},
  eprintclass = {hep-th},
  eprinttype  = {arXiv},
}

@Article{CCFR18,
  author      = {Caroca, R. and Concha, P. and Fierro, O. and Rodr{\'i}guez, E.},
  journal     = {Phys. Lett. B},
  title       = {Three-dimensional Poincaré supergravity and $N$-extended supersymmetric BMS${}_3$ algebra},
  year        = {2019},
  pages       = {93},
  volume      = {792},
  eprint      = {1812.05065},
  eprintclass = {hep-th},
  eprinttype  = {arXiv},
}

@Article{BBLN18,
  author      = {Banerjee, N. and Bhattacharjee, A. and Lodato, I. and Neogi, T.},
  journal     = {JHEP},
  title       = {Maximally $N$-extended super-BMS${}_3$ algebras and generalized 3D gravity solutions},
  year        = {2019},
  pages       = {115},
  volume      = {01},
  eprint      = {1807.06768},
  eprintclass = {hep-th},
  eprinttype  = {arXiv},
}

@Article{BDR17,
  author      = {Basu, R. and Detournay, S. and Riegler, M.},
  journal     = {JHEP},
  title       = {Spectral flow in 3D flat spacetimes},
  year        = {2017},
  pages       = {134},
  volume      = {12},
  eprint      = {1706.07438},
  eprintclass = {hep-th},
  eprinttype  = {arXiv},
}

@Article{FMT17,
  author      = {Fuentealba, O. and Matulich, J. and Troncoso, R.},
  journal     = {JHEP},
  title       = {Asymptotic structure of $N=2$ supergravity in 3D: extended super-BMS${}_{3}$ and nonlinear energy bounds},
  year        = {2017},
  pages       = {030},
  volume      = {09},
  eprint      = {1706.07542},
  eprintclass = {hep-th},
  eprinttype  = {arXiv},
}

@Article{BLN17,
  author      = {Banerjee, N. and Lodato, I. and Neogi, T.},
  journal     = {Phys. Rev. D},
  title       = {$N=4$ Supersymmetric BMS3 algebras from asymptotic symmetry analysis},
  year        = {2017},
  pages       = {066029},
  volume      = {96},
  eprint      = {1706.02922},
  eprintclass = {hep-th},
  eprinttype  = {arXiv},
}

@Article{LM16,
  author      = {Lodato, I. and Merbis, W.},
  journal     = {JHEP},
  title       = {Super-BMS${}_3$ algebras from $N=2$ flat supergravities},
  year        = {2016},
  pages       = {150},
  volume      = {11},
  eprint      = {1610.07506},
  eprintclass = {hep-th},
  eprinttype  = {arXiv},
}

@Article{BJLMN16,
  author      = {Banerjee, N. and Jatkar, D.P. and Lodato, I. and Mukhi, S. and Neogi, T.},
  journal     = {JHEP},
  title       = {Extended supersymmetric BMS${}_3$ algebras and their free field realisations},
  year        = {2016},
  pages       = {1--23},
  volume      = {11},
  eprint      = {1609.09210},
  eprintclass = {hep-th},
  eprinttype  = {arXiv},
}

@Article{BG09,
  author      = {Bagchi, A. and Gopakumar, R.},
  journal     = {JHEP},
  title       = {Galilean conformal algebras and AdS/CFT},
  year        = {2009},
  pages       = {037},
  volume      = {07},
  eprint      = {0902.1385},
  eprintclass = {hep-th},
  eprinttype  = {arXiv},
}

@Article{BGMM10,
  author      = {Bagchi, A. and Gopakumar, R. and Mandal, I. and Miwa, A.},
  journal     = {JHEP},
  title       = {GCA in 2D},
  year        = {2010},
  pages       = {004},
  volume      = {08},
  eprint      = {0912.1090},
  eprintclass = {hep-th},
  eprinttype  = {arXiv},
}

@Article{GG11,
  author      = {Gaberdiel, M.R. and Gopakumar, R.},
  journal     = {Phys. Rev. D},
  title       = {An $AdS_3$ dual for minimal model CFTs},
  year        = {2011},
  pages       = {066007},
  volume      = {83},
  eprint      = {1011.2986},
  eprintclass = {hep-th},
  eprinttype  = {arXiv},
}

@Article{Vas12,
  author      = {Vasiliev, M.A.},
  journal     = {J. Phys. A},
  title       = {Holography, unfolding and higher-spin theory},
  year        = {2013},
  pages       = {21},
  volume      = {46},
  eprint      = {1203.5554},
  eprintclass = {hep-th},
  eprinttype  = {arXiv},
}

@Article{ABFGR13,
  author      = {Afshar, H. and Bagchi, A. and Fareghbal, R. and Grumiller, D. and Rosseel, J.},
  journal     = {Phys. Rev. Lett.},
  title       = {Higher spin theory in 3-dimensional flat space},
  year        = {2013},
  pages       = {121603},
  volume      = {111},
  eprint      = {1307.4768},
  eprintclass = {hep-th},
  eprinttype  = {arXiv},
}

@Article{BO14,
  author      = {Barnich, G. and Oblak, B.},
  journal     = {JHEP},
  title       = {Notes on the BMS group in three dimensions: I. Induced representations},
  year        = {2014},
  pages       = {129},
  volume      = {06},
  eprint      = {1403.5803},
  eprintclass = {hep-th},
  eprinttype  = {arXiv},
}

@Book{Obl17,
  author    = {Oblak, B.},
  publisher = {Springer International Publishing},
  title     = {BMS particles in three dimensions},
  year      = {2017},
}

@Article{Bag13,
  author      = {Bagchi, A.},
  journal     = {JHEP},
  title       = {Tensionless strings and Galilean conformal algebra},
  year        = {2013},
  pages       = {141},
  volume      = {05},
  eprint      = {1303.0291},
  eprintclass = {hep-th},
  eprinttype  = {arXiv},
}

@Article{BBCP18,
  author      = {Bagchi, A. and Banerjee, A. and Chakrabortty, S. and Parekh, P.},
  journal     = {JHEP},
  title       = {Inhomogeneous tensionless superstrings},
  year        = {2018},
  pages       = {065},
  volume      = {02},
  eprint      = {1710.03482},
  eprintclass = {hep-th},
  eprinttype  = {arXiv},
}

@Article{BBCP20,
  author      = {Bagchi, A. and Banerjee, A. and Chakrabortty, S. and Parekh, P.},
  journal     = {Nucl. Phys. B},
  title       = {Exotic origins of tensionless superstrings},
  year        = {2020},
  pages       = {135139},
  volume      = {801},
  eprint      = {1811.10877},
  eprintclass = {hep-th},
  eprinttype  = {arXiv},
}

@Article{BBCDP20,
  author      = {Bagchi, A. and Banerjee, A. and Chakrabortty, S. and Dutta, S. and Parekh, P.},
  journal     = {JHEP},
  title       = {A tale of three. Tensionless strings and vacuum structure},
  year        = {2020},
  pages       = {061},
  volume      = {04},
  eprint      = {2001.00354},
  eprintclass = {hep-th},
  eprinttype  = {arXiv},
}

@Article{Hag72,
  author  = {Hagen, C.R.},
  journal = {Phys. Rev. D},
  title   = {Scale and conformal transformations in Galilean-covariant field theory},
  year    = {1972},
  pages   = {377--388},
  volume  = {5},
}

@Article{DH09,
  author      = {Duval, C. and Horv{\'a}thy, P.A.},
  journal     = {J. Phys. A},
  title       = {Non-relativistic conformal symmetries and Newton-Cartan structures},
  year        = {2009},
  pages       = {465206},
  volume      = {42},
  eprint      = {0904.0531},
  eprintclass = {math-ph},
  eprinttype  = {arXiv},
}

@Article{BM10,
  author      = {Bagchi, A. and Mandal, I.},
  journal     = {Phys. Rev. D},
  title       = {Supersymmetric extension of Galilean conformal algebras},
  year        = {2010},
  pages       = {8--15},
  volume      = {80},
  eprint      = {0905.0580},
  eprintclass = {hep-th},
  eprinttype  = {arXiv},
}

@Article{HR10,
  author      = {Hosseiny, A. and Rouhani, S.},
  journal     = {J. Math. Phys.},
  title       = {Affine extension of Galilean conformal algebra in $2 + 1$ dimensions},
  year        = {2010},
  pages       = {052307},
  volume      = {51},
  eprint      = {0909.1203},
  eprintclass = {hep-th},
  eprinttype  = {arXiv},
}

@Article{IW53,
  author  = {In{\"o}n{\"u}, E. and Wigner, E.P.},
  journal = {Proc. Nat. Acad. Sci.},
  title   = {On the contraction of groups and their representations},
  year    = {1953},
  pages   = {510--524},
  volume  = {39},
}

@Article{Sale61,
  author  = {Saletan, E.J.},
  journal = {J. Math. Phys.},
  title   = {Contraction of Lie groups},
  year    = {1961},
  pages   = {1--21},
  volume  = {2},
}

@Article{DK03,
  author  = {D'Appollonio, G. and Kiritsis, E.},
  journal = {Nucl. Phys. B},
  title   = {String interactions in gravitational wave backgrounds},
  year    = {2003},
  pages   = {80--170},
  volume  = {674},
  eprint      = {hep-th/0305081},
  eprinttype  = {arXiv},
}

@Article{BDKZ04,
  author  = {Bianchi, M. and D'Appollonio, G. and Kiritsis, E. and Zapata, O.},
  journal = {JHEP},
  title   = {String amplitudes in the Hpp-wave limit of $AdS_3 \times S^3$},
  year    = {2004},
  pages   = {074},
  volume  = {04},
  eprint      = {hep-th/0402004},
  eprinttype  = {arXiv},
}

@Article{Tak71,
  author  = {Takiff, S.},
  journal = {Trans. Amer. Math. Soc.},
  title   = {Rings of invariant polynomials for a class of Lie algebras},
  year    = {1971},
  pages   = {249--262},
  volume  = {160},
}

@Article{BR12,
  author      = {Babichenko, A. and Ridout, D.},
  journal     = {J. Phys. A},
  title       = {Takiff superalgebras and conformal field theory},
  year        = {2013},
  pages       = {125204},
  volume      = {46},
  eprint      = {1210.7094},
  eprintclass = {math-ph},
  eprinttype  = {arXiv},
}

@Article{Que20,
  author      = {Quella, T.},
  journal     = {J. Phys. Commun.},
  title       = {On conformal field theories based on Takiff superalgebras},
  year        = {2020},
  pages       = {075013},
  volume      = {4},
  eprint      = {2004.06456},
  eprintclass = {hep-th},
  eprinttype  = {arXiv},
}

@Article{DQ08,
  author      = {D'Appollonio, G. and Quella, T.},
  journal     = {JHEP},
  title       = {The diagonal cosets of the Heisenberg group},
  year        = {2008},
  pages       = {060},
  volume      = {05},
  eprint      = {0801.4634},
  eprintclass = {hep-th},
  eprinttype  = {arXiv},
}

@Article{Zam85,
  author      = {Zamolodchikov, A.B.},
  journal     = {Theor. Math. Phys.},
  title       = {Infinite additional symmetries in two-dimensional conformal quantum field theory},
  year        = {1985},
  pages       = {1205--1213},
  volume      = {65},
}

@Article{CCRS18,
  author      = {Caroca, R. and Concha, P. and Rodr{\'i}guez, E. and Saldago-Rebolledo, P.},
  journal     = {Eur. Phys. J. C},
  title       = {Generalizing the $\mathfrak{bms}_3$ and $2$D-conformal algebras by expanding the Virasoro algebra},
  year        = {2018},
  pages       = {262},
  volume      = {78},
  eprint      = {1707.07209},
  eprintclass = {hep-th},
  eprinttype  = {arXiv},
}

@Online{CIRR20,
  author      = {Concha, P. and Ipinza, M. and Ravera, L. and Rodr{\'i}guez, E.},
  eprint      = {2010.01216},
  eprintclass = {hep-th},
  eprinttype  = {arXiv},
  title       = {Non-relativistic three-dimensional supergravity theories and semigroup expansion method},
}

@Article{BH86,
  author  = {Brown, J.D. and Henneaux, M.},
  journal = {Commun. Math. Phys.},
  title   = {Central charges in the canonical realization of asymptotic symmetries: An example from three-dimensional gravity},
  year    = {1986},
  pages   = {207--226},
  volume  = {104},
}

@Article{BC07,
  author     = {Barnich, G. and Comp{\'e}re, G.},
  journal    = {Class. Quant. Grav.},
  title      = {Classical central extension for asymptotic symmetries at null infinity in three spacetime dimensions},
  year       = {2007},
  pages      = {F15--F23},
  volume     = {24},
  eprint     = {gr-qc/0610130},
  eprinttype = {arXiv},
}

@Article{BL13,
  author      = {Barnich, G. and Lambert, P.-H.},
  journal     = {Rom. J. Phys.},
  title       = {Asymptotic symmetries at null infinity and local conformal properties of spin coefficients},
  year        = {2013},
  pages       = {5},
  volume      = {58},
  eprint      = {1301.5754},
  eprintclass = {gr-qc},
  eprinttype  = {arXiv},
}

@Article{GMPT13,
  author      = {Gonz{\'a}lez, H.A. and Matulich, J. and Pino, M. and Troncos, R.},
  journal     = {JHEP},
  title       = {Asymptotically flat space times in three-dimensional higher spin gravity},
  year        = {2013},
  pages       = {016},
  volume      = {09},
  eprint      = {1307.5651},
  eprintclass = {hep-th},
  eprinttype  = {arXiv},
}

@Article{CF17,
  author      = {Comp{\'e}re, G. and Fiorucci, A.},
  journal     = {Class. Quant. Grav.},
  title       = {Asymptotically flat spacetimes with $BMS_{3}$ symmetry},
  year        = {2017},
  pages       = {20},
  volume      = {34},
  eprint      = {1705.06217},
  eprintclass = {hep-th},
  eprinttype  = {arXiv},
}

@Article{GRR14,
  author      = {Grumiller, D. and Riegler, M. and Rosseel, J.},
  journal     = {JHEP},
  title       = {Unitarity in three-dimensional flat space higher spin theories},
  year        = {2014},
  pages       = {015},
  volume      = {07},
  eprint      = {1403.5297},
  eprintclass = {hep-th},
  eprinttype  = {arXiv},
}

@Article{RR17,
  author      = {Rasmussen, J. and Raymond, C.},
  journal     = {Nucl. Phys. B},
  title       = {Galilean contractions of $W$-algebras},
  year        = {2017},
  pages       = {435--479},
  volume      = {922},
  eprint      = {1701.04437},
  eprintclass = {hep-th},
  eprinttype  = {arXiv},
}

@PhdThesis{Thielemans,
  author = {Thielemans, K.},
  school = {Katholieke Universiteit Leuven},
  title  = {An Algorithmic Approach to Operator Product Expansions, W-Algebras and W-Strings},
  year   = {1994},
}

@Online{DZ07,
  author      = {Zhang, W. and Dong, C.},
  eprint      = {0711.4624},
  eprintclass = {math.QA},
  eprinttype  = {arXiv},
  title       = {W-algebra $W(2,2)$ and the vertex operator algebra $L(1/2,0)\otimes L(1/2,0)$},
}

@Article{RR19,
  author      = {Rasmussen, J. and Raymond, C.},
  journal     = {Nucl. Phys. B},
  title       = {Higher-order Galilean contractions},
  year        = {2019},
  pages       = {114680},
  volume      = {945},
  eprint      = {1901.06069},
  eprintclass = {hep-th},
  eprinttype  = {arXiv},
}

@Article{RRR20,
  author      = {Ragoucy, E. and Rasmussen, J. and Raymond, C.},
  journal     = {Nucl. Phys. B},
  title       = {Multi-graded Galilean conformal algebras},
  year        = {2020},
  pages       = {115092},
  volume      = {957},
  eprint      = {2002.08637},
  eprintclass = {hep-th},
  eprinttype  = {arXiv},
}

@Article{Moham94,
  author     = {Mohammedi, N.},
  journal    = {Phys. Lett. B},
  title      = {On bosonic and supersymmetric current algebras for non-semisimple groups},
  year       = {1994},
  pages      = {371--376},
  volume     = {325},
  eprint     = {hep-th/9312182},
  eprinttype = {arXiv},
}

@Article{ORS94,
  author     = {Olive, D.I. and Rabinovici, E. and Schwimmer, A.},
  journal    = {Phys. Lett. B},
  title      = {A class of string backgrounds as a semiclassical limit of WZW models},
  year       = {1994},
  pages      = {361--364},
  volume     = {321},
  eprint     = {hep-th/9311081},
  eprinttype = {arXiv},
}

@Article{Sfet94,
  author     = {Sfetsos, K.},
  journal    = {Phys. Rev. D},
  title      = {Gauged WZW models and non-abelian duality},
  year       = {1994},
  pages      = {2784--2798},
  volume     = {50},
  eprint     = {hep-th/9402031},
  eprinttype = {arXiv},
}

@Article{HKOC96,
  author     = {Halpern, M.B. and Kiritsis, E. and Obers, N. and Clubok, K.},
  journal    = {Phys. Rep.},
  title      = {Irrational conformal field theory},
  year       = {1996},
  pages      = {1--138},
  volume     = {265},
  eprint     = {hep-th/9501144},
  eprinttype = {arXiv},
}

@Article{FS94,
  author     = {Figueroa-O'Farrill, J.M. and Stanciu, S.},
  journal    = {Phys. Lett. B},
  title      = {Nonsemisimple Sugawara constructions},
  year       = {1994},
  pages      = {40--46},
  volume     = {327},
  eprint     = {hep-th/9402035},
  eprinttype = {arXiv},
}

@Article{NW93,
  author     = {Nappi, C.R. and Witten, E.},
  journal    = {Phys. Rev. Lett.},
  title      = {A WZW model based on a non-semi-simple group},
  year       = {1993},
  pages      = {3751--3753},
  volume     = {71},
  eprint     = {hep-th/9310112},
  eprinttype = {arXiv},
}

@Article{HK89,
  author     = {Halpern, M.B. and Kiritsis, E.},
  journal    = {Mod. Phys. Lett. A},
  title      = {General Virasoro construction on affine $\mathfrak{g}$},
  year       = {1989},
  pages      = {1373--1380},
  volume     = {4},
}

@Article{KK93,
  author     = {Kiritsis, E. and Kounnas, C.},
  journal    = {Phys. Lett. B},
  title      = {String propagation in gravitational wave backgrounds},
  year       = {1994},
  pages      = {264--272},
  volume     = {320},
  eprint     = {hep-th/9310202},
  eprinttype = {arXiv},
}

@Online{BJP11,
  author      = {Bao, Y. and Jiang, C. and Pei, Y.},
  eprint      = {1104.3921},
  eprintclass = {math.QA},
  eprinttype  = {arXiv},
  title       = {Representations of affine Nappi-Witten algebras},
}

@Article{BF12,
  author      = {Bagchi, A. and Fareghbal, R.},
  journal     = {JHEP},
  title       = {BMS/GCA redux: Towards flatspace holography from non-relativistic symmetries},
  year        = {2012},
  pages       = {092},
  volume      = {10},
  eprint      = {1203.5795},
  eprintclass = {hep-th},
  eprinttype  = {arXiv},
}

@Online{BKRS20,
  author      = {Babichenko, A. and Kawasetsu, K. and Ridout, D. and Stewart, W.},
  eprint      = {2011.14453},
  eprintclass = {math-ph},
  eprinttype  = {arXiv},
  title       = {Representations of the Nappi-Witten vertex operator algebra},
}

\end{document}